%% file: paper.tex
\documentclass[letterpaper,twocolumn,10pt]{article}
\usepackage{usenix-2020-09}

\usepackage{amsmath}
\usepackage{amssymb}
\usepackage{xspace}
\usepackage{enumitem}
\usepackage{graphicx}
\usepackage{soul}
\usepackage{adjustbox}
\usepackage{inconsolata}
\usepackage{booktabs}
\usepackage{listings}
\usepackage{color}
\usepackage{colortbl}
\usepackage{multirow}
\usepackage{subcaption}
\usepackage{authblk}

\DeclareMathOperator*{\argmax}{arg\,max}  

\makeatletter
\newcommand\notsotiny{\@setfontsize\notsotiny\@vipt\@viipt}
\makeatother

\definecolor{codegreen}{rgb}{0,0.6,0}
\definecolor{codeblue}{rgb}{0,0.5,1.0}
\definecolor{codegray}{rgb}{0.5,0.5,0.5}
\definecolor{codepurple}{rgb}{0.58,0,0.82}
\definecolor{backcolour}{rgb}{0.95,0.95,0.92}

\lstdefinestyle{mystyle}{
    commentstyle=\color{codeblue},
    keywordstyle=\bfseries\color{black},
    numberstyle=\tiny\color{codegray},
    stringstyle=\color{codepurple},
    basicstyle=\ttfamily\footnotesize,
    breakatwhitespace=false,
    breaklines=true,
    captionpos=b,
    keepspaces=true,
    numbers=left,
    numbersep=5pt,
    showspaces=false,
    showstringspaces=false,
    showtabs=false,
    tabsize=2,
    language=C,
    otherkeywords={get_API_endpoint, upload_model, define_model_group, activate_model_group, retire_model, retire_model_group, get_model_groups, request_inf, verify_cert}, 
}

\usepackage{subcaption}

\usepackage[font={small, bf, sf, stretch=0.9}, textfont={small, sf}]{caption}
\usepackage{cleveref}
\crefname{algocf}{alg.}{algs.}
\Crefname{algocf}{Alg.}{Algs.}
\Crefname{equation}{Eq.}{Eqs.}
\Crefname{figure}{Fig.}{Figs.}
\Crefname{table}{Tab.}{Tabs.}

\usepackage[subtle,wordspacing=normal]{savetrees} 

\usepackage[linesnumbered,ruled,nofillcomment, norelsize]{algorithm2e}
\SetAlFnt{\footnotesize}
\SetKwRepeat{Do}{do}{while}%
\SetKwFor{While}{while}{}{end while}%
\SetKwBlock{KwFunction}{function}{}
\SetKwBlock{KwOn}{on}{}
\SetKwBlock{KwRepeat}{repeat}{}
\SetKwBlock{KwPeriodically}{periodically}{}
\SetKwBlock{KwState}{state}{}
\SetKwBlock{KwAtomic}{atomically}{}
\SetKwBlock{KwConcurrently}{concurrently}{}
\SetCommentSty{textup} 
\SetKwComment{KwIttayComment}{(}{)} 
\SetKw{KwAnd}{and} 
\SetKw{KwOr}{or}
\SetKw{KwNot}{not}
\SetKw{KwSetTimer}{setTimer}
\SetKwBlock{KwExpTimer}{on timer expiration}{}
\SetKwFor{ForAll}{forall}{}

\SetKwFunction{FMain}{Main}
\SetKwProg{Fn}{fn}{}{}
\SetKwInput{KwPre}{Pre}

\usepackage{amsthm,lipsum}
\newtheoremstyle{theorem-small}
{3pt}
{3pt}
{\itshape}
{}
{\bfseries}
{.}
{.5em}
{}
\theoremstyle{theorem-small}
\newtheorem{theorem-small}{Theorem}[section]

\newenvironment{proof-sketch}{\noindent\textit{Proof (sketch).}}{\hfill}

\newlength{\gapspace}

\newcommand{\unit}[1]{\mbox{\hspace{2pt}#1}\xspace}

\usepackage{filecontents}

\definecolor{AlexBlue}{HTML}{0048BA}
\definecolor{PrpOrange}{HTML}{ff8c00}

\usepackage[normalem]{ulem}

\newenvironment{myitemize}{%
\begin{itemize}[leftmargin=1em, itemsep=.1em, parsep=.1em, topsep=.1em,
    partopsep=.1em]}
{\end{itemize}}

\newenvironment{myenumerate}{%
\begin{enumerate}[leftmargin=1.50em, itemsep=0em, parsep=0em, topsep=.1em,
    partopsep=.1em]}
{\end{enumerate}}

\newenvironment{structure*}{\color{blue}\begin{myenumerate}}{\end{myenumerate}}

\usepackage{newfloat}
\DeclareFloatingEnvironment[fileext=lop]{listing}

\newcommand{\System}{\sys}
\newcommand{\sys}{Dropbear\xspace}

\newcommand{\infproxy}{inference proxy\xspace}

\newcommand{\tinyskip}{\vspace{3pt}}
\newcommand{\mypar}[1]{\tinyskip\noindent\textbf{#1.}\xspace}
\newcommand{\myparr}[1]{\tinyskip\noindent\textbf{#1}\xspace}
\newcommand{\mypari}[1]{\tinyskip\noindent\emph{#1.}\xspace}
\newcommand{\myparii}[1]{\tinyskip\noindent\emph{#1}\xspace}

\usepackage[compact]{titlesec}

\newcommand{\msg}[1]{\textsf{\small{}#1}\xspace}
\newcommand{\eg}{e.g.,\xspace}
\newcommand{\ie}{i.e.,\xspace}
\newcommand{\etal}{et al.\xspace}

\newcommand{\F}{\mbox{Fig.\hspace{0.25em}}}
\newcommand{\T}{\mbox{Tab.\hspace{0.25em}}}

\usepackage{tikz}
\newcommand*\myc[1]{%
\scalebox{0.78}{\begin{tikzpicture}[baseline=-4pt]
  \node[draw,circle,inner sep=0.5pt, fill=black] {\textcolor{white}{\textsf{\textbf{#1}}}};
\end{tikzpicture}}}

\begin{document}

\date{}

\title{\Large \bf \sys: Machine Learning Marketplaces made Trustworthy\\ with Byzantine Model Agreement}
\author[1,2]{Alex Shamis}
\author[1,2]{Peter Pietzuch}
\author[1]{Antoine Delignat-Lavaud}
\author[1,3]{Andrew Paverd}
\author[1]{Manuel Costa}
\affil[1]{Microsoft Research}
\affil[2]{Imperial College London}
\affil[3]{Microsoft Security Response Center}

\maketitle

\input{0-abstract.tex}
\input{1-introduction.tex}
\input{2-byzantine-inference.tex}

\input{3-overview.tex}
\input{5-design.tex}
\input{6-evaluation.tex}
\input{7-related-work.tex}
\input{8-conclusion.tex}

\bibliographystyle{acm}
\bibliography{paper}

\end{document}

%% file: 0-abstract.tex
\begin{abstract}
  Marketplaces for machine learning~(ML) models are emerging as a way for organizations to monetize models. They allow model owners to retain control over hosted models by using cloud resources to execute ML inference requests for a fee, preserving model confidentiality. Clients that rely on hosted models require trustworthy inference results, even when models are managed by third parties. While the resilience and robustness of inference results can be improved by combining multiple independent models, such support is unavailable in today's marketplaces. 

  We describe \emph{\sys}, the first ML model marketplace that provides clients with strong integrity guarantees by combining results from multiple models in a trustworthy fashion. \sys replicates inference computation across a \emph{model group}, which consists of multiple cloud-based GPU nodes belonging to different model owners. Clients receive \emph{inference certificates} that prove agreement using a Byzantine consensus protocol, even under model heterogeneity and concurrent model updates. To improve performance, \sys batches inference and consensus operations separately: it first performs the inference computation across a model group, before ordering requests and model updates. Despite its strong integrity guarantees, \sys{}'s performance matches that of state-of-the-art ML inference systems: deployed across 3~cloud sites, it handles 800\unit{requests/s} with ImageNet models.
\end{abstract}

%% file: 1-introduction.tex
\section{Introduction}

Many applications use inference computation over machine learning~(ML) models, particularly deep neural networks~(DNNs), and their number is only expected to grow~\cite{GartnerI61:online, natale2020imagining, krittanawong2018rise, khakurel2018rise}. This gives rise to ML marketplaces such as AWS Marketplace~\cite{catania_keefer_1987}, Stream.ML~\cite{StreamML22:online}, Modzy~\cite{AIModelM99:online}, and ModelPlace.ai~\cite{Modelpla60:online}, enabling organizations to monetize ML models~\cite{Roboflow81:online, OutOfThe46:online, BigVisio20:online}. ML marketplaces host ML models, which clients can use to execute inference requests for a fee. They are typically cloud-based, using cloud nodes with GPUs or other AI accelerators to execute inference requests.

The rising popularity of ML marketplaces comes from their combination of features. They offer organisations the ability to retain control over their ML models, \eg serving inference requests without revealing the internals of the models to clients. They expose simple APIs~\cite{AzureMac41:online, AmazonSa44:online} that allow organizations to upload models and clients to submit inference requests against these models. Brokerage or search functionality~\cite{Sommelier, romero2021infaas, crankshaw2017clipper} helps clients discover relevant models for a given inference application.

In many applications that use ML inference \eg for decision-making, the correctness of inference results affects the trustworthiness of the application. For example, ML inference is used by the insurance industry to decide if an insurance application is underwritten~\cite{hall2017artificial, aggour2006automating, maier2019transforming}, or by courts to aid in sentencing within the criminal justice system~\cite{coglianese2020ai, hillman2019use, kugler2018ai}. Existing ML marketplaces, however, require clients submitting inference requests to fully trust the model owners. A model may have been poisoned during training, or a malicious or incompetent model owner may tamper with inference results to save on computational resources (resulting in lost accuracy) or alter the behaviour of an application that depends on inference results.

Therefore, it is challenging for clients to establish the trustworthiness of model owners and the inference results for specific models~\cite{wan2022automated, jameel2015automatic, gao2019ai}. For example, if a client wants to establish correctness, they must redo the inference computation to reproduce the result, but model owners are reluctant to release models for confidentiality reasons. In addition, the numeric values of inference results depend on the model version, the employed hardware (\eg the GPU type~\cite{whitehead2011precision}), and the specific software stack~\cite{chen2018tvm, cyphers2018intel, rotem2018glow, sabne2020xla}, making reproducibility of results challenging.

A promising approach for increasing the robustness and resilience of inference results is to \emph{aggregate} results from multiple independent models~\cite{jia2020intrinsic, biggio2011bagging, smutz2016tree}. Prior work has shown how such model aggregation can mitigate poisoning attacks, which corrupt the integrity of the model at training time unbeknownst to clients~\cite{liu2017trojaning,koh2022stronger, mei2015using}, and evasion attacks, in which carefully crafted adversarial inference requests can lead to incorrect results~\cite{dalvi2004adversarial, szegedy2013intriguing, eykholt2018robust}. To protect against a subset of models being malicious or having been created using a poisoned dataset, model aggregation techniques require models to be created by independent model owners on independent datasets. While ML marketplaces support functionality for clients to discover relevant models~\cite{Sommelier, romero2021infaas, crankshaw2017clipper}, they lack systems support for trustworthy model aggregation.

We describe \textbf{\sys}, a cloud-based trustworthy ML marketplace that supports inference decisions over models through Byzantine model agreement. \sys allows model owners to upload models, which are grouped into \emph{model groups} of related models. Each model group accepts the same type of inference request and its models produce equivalent predictions, allowing them to verify one another's inference results. Clients submit inference requests against model groups, which are executed using Byzantine agreement by GPU nodes under the control of the model owners. The technical contributions of \sys are as follows:

\tinyskip

\noindent
(1)~\sys realizes \textbf{Byzantine model agreement} in which inference results from potentially dishonest model owners must reach agreement. Models are hosted on cloud nodes controlled by the model owners to safeguard their models. Inference computation and model updates are replicated across multiple nodes, which must reach agreement on inference results. To support combining results from a group of independent models, agreement is defined subject to acceptable bounds on the results. This limits possible divergence if a dishonest node provides incorrect results. Consensus among the nodes regarding model versions, the membership in a model group, and inference results is reached using a Byzantine agreement protocol~\cite{pbft, yin2018hotstuff, bessani2014state}.

\tinyskip

\noindent
(2)~\sys follows an \textbf{execute-agree-attest} strategy, which improves performance by separating the expensive execution of inference requests from the agreement and attestation of inference results: (i)~nodes first \emph{execute} inference requests as part of \emph{execution batches}; (ii)~after execution, a primary node batches the inference requests again in \emph{consensus batches}. It orders them with respect to model updates, and sends batches to a set of backup nodes. The nodes then \emph{agree} on an inference result based on the model group membership and its bounds; and (iii)~after $2/3$ of the nodes have agreed, they \emph{attest} that the result is within the prescribed bounds.

By batching inference requests and consensus operations independently, \sys allows for a higher degree of parallelism across nodes: execution batches for heavily-referenced models~\cite{jouppi2017datacenter} increase GPU utilisation; consensus batches are constructed to better utilize the network between nodes and reduce cryptographic overhead.

\tinyskip

\noindent
(3)~Despite supporting Byzantine model agreement, \sys maintains an API that remains compatible with current cloud-based ML marketplace APIs by having only a single API endpoint~\cite{AzureMac41:online, AmazonSa44:online, CloudInf6:online}. The complexity of model agreement is hidden from clients: clients receive inference results together with an \textbf{inference certificate}, which is a universally-verifiable proof that the nodes that executed the models from the model group agreed on the result.

To offer clients a single API endpoint, the inference certificate is created by an \emph{\infproxy}, which combines cryptographically-signed Merkle trees of inference results and consensus messages. The \infproxy is untrusted: clients check the inference certificate's correctness by verifying that its content is signed by the service and that it contains the client's inference request.

\tinyskip

\noindent
Our implementation of \sys is written in 5,000~lines of C++. It supports DNN models defined in the ONNX format~\cite{onnx} and served by the ONNX Runtime~\cite{onnx_runtime}, a cross-platform inference library implementation with heterogeneous accelerator support. Our experiments using the Imagenet~\cite{deng2009imagenet} models from the ONNX and PyTorch model zoos~\cite{GitHubon40:online, 10ModelZ20:online} show that \sys can provide inference certificates without compromising performance: \sys handles close to 800~inference requests/s when using 16~cloud nodes with a total of 32 NVIDIA V100 GPUs with 16~\unit{GB} of memory, distributed across 3 cloud sites.


%% file: 2-byzantine-inference.tex
\section{Trust in ML Marketplaces}
\label{sec:background}

We introduce cloud-based ML marketplaces (\S\ref{sec:background:inference_service}), describe their security challenges~({\S\ref{subsec:motivation}}) and describe our assumed threat model~(\S\ref{subsec:threat-model}).

\subsection{ML marketplaces}
\label{sec:background:inference_service}

All major cloud providers offer ML services that host pre-trained models, \eg AWS Marketplace~\cite{catania_keefer_1987}, Azure Cognitive Services~\cite{cognitive_services}, and Google Cloud AI APIs~\cite{google_ai_ml}. As a representative example, AWS Marketplace is a \emph{ML marketplace} in which users can upload models or pay for the execution of inference requests against hosted models. Other marketplaces, including Stream.ML~\cite{StreamML22:online}, Modzy~\cite{AIModelM99:online}, and ModelPlace.ai~\cite{Modelpla60:online}, focus on specific application domains or business models. Clients use models from ML marketplaces to take advantage of pre-trained models, avoid having to build up ML expertise in-house, and reduce costs~\cite{kumar2020marketplace, httpsass80:online, AIPricin97:online}.


Many of these models are trained using deep learning, which is a training technique for deep neural network~(DNN) models~\cite{Goodfellow-et-al-2016}. Such models have shown great success in diverse areas, including image classification, object discovery, natural language processing, and fraud detection. Training DNN models is often computationally expensive and typically requires a large amount of training data, which may be difficult to acquire. Achieving high accuracy also requires advanced ML expertise, \eg in order to tune the model's hyper-parameters. As such, ML marketplaces allow organizations to use accurate DNN models with no upfront cost.



\begin{figure}[tb]
  \centering
  \includegraphics[width=.9\columnwidth]{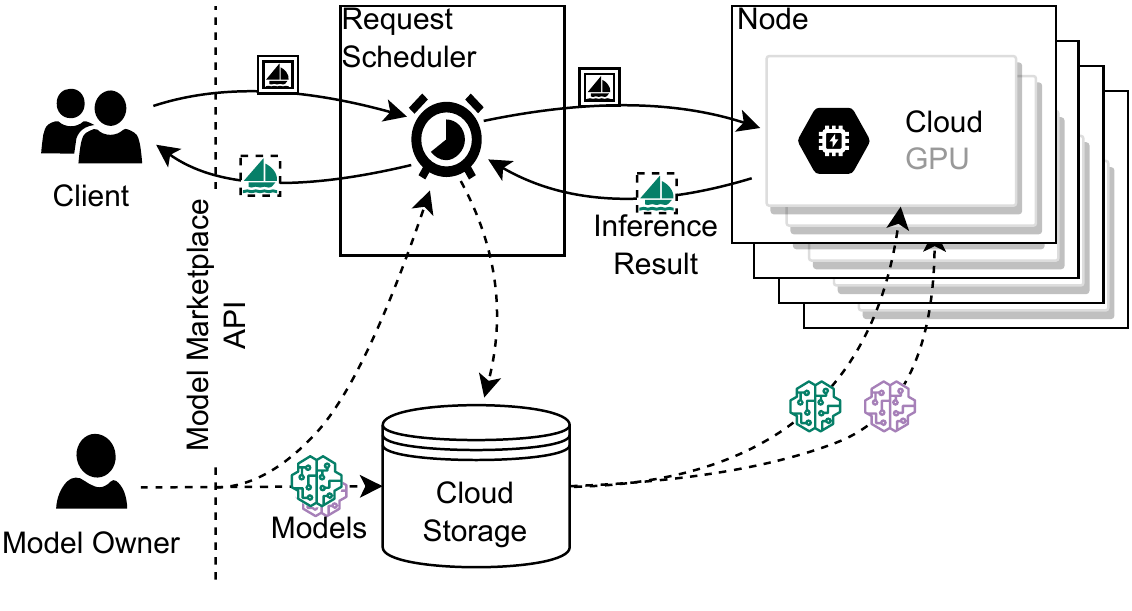}
  \caption{ML marketplace \textnormal{(A model owner uploads an ML model to cloud storage and hosts its execution on a set of cloud nodes. A client then sends an inference request of a sail boat, which the marketplace routes to a corresponding model before returning an inference result.)}}
  \label{fig:INFaas}
\end{figure}

\Cref{fig:INFaas} gives an overview of a cloud-based ML marketplace~\cite{kim2014inference}. It exposes an API that is used by \emph{model owners} and \emph{clients}: model owners upload one or more models, which are typically expressed in a standardized format, such as ONNX~\cite{onnx} or NNEF~\cite{debreczeni2019neural}. The models are deployed on \emph{cloud nodes} with hardware accelerators such as GPUs. The nodes execute inference requests submitted by clients using an \emph{inference engine} implementation.

Clients submit inference requests through an application protocol, such as HTTP, and pay for each submitted request. An inference request specifies a model (or a set of compatible models) and the input data for the inference request, \eg an image. The request is then forwarded to a cloud node's inference engine for execution.

\subsection{Trust requirements}
\label{subsec:motivation}


When clients use ML inference as part of an application, the application's trustworthiness depends on the correct execution of inference requests. Across different application domains, ML marketplaces present a unique challenge for clients trusting inference results when the computation occurs on cloud nodes using ML models created and controlled by third parties:


\mypar{Content moderation} The growth in user-created content on the internet~\cite{Usergen45:online, YouTube12:online, mivsura2016data} puts an increasing amount of pressure on online services to moderate public content, \eg user forum posts, customer reviews, and shared images and videos. Automated ML-based approaches for content moderation have been widely adopted~\cite{MachineL63:online, gorwa2020algorithmic, gillespie2020content}, but untrustworthy moderation due to flawed or compromised ML models can result in financial losses or legal consequences~\cite{TheHidde96:online, ContentM17:online, Contentm50:online}.

\mypar{Automated decision-making} ML has helped remove or reduce manual tasks of employees. For example, issuing insurance policies can be automated by using ML models, with initial real-world deployments~\cite{hall2017artificial, aggour2006automating, maier2019transforming}; decision-making about mortgages has been handled similarly~\cite{Whatisth51:online}. Such automated decision-making, however, has real-world financial implications, and thus creates incentives for fraud and bias~\cite{Amazonsc3:online}.

\mypar{Anomaly detection} For automated anomaly detection, ML models process large amounts of data to find anomalous datapoints~\cite{nassif2021machine, elmrabit2020evaluation, choi2018generative}. Anomaly detection is used in scenarios ranging from the discovery of attacks on network infrastructure to identifying fraudulent financial transactions~\cite{marchal2019detecting}. Such scenarios typically carry a high cost of false negatives, and adversaries have an incentive to circumvent anomaly detection \eg by poisoning the employed ML model.


\mypar{Supply chains} Manufacturing and retail organizations manage supply chains in order to reduce the time between an inventory item's delivery and consumption, while ensuring that items are always available. ML techniques have been used in supply chain optimization~\cite{wenzel2019literature, carbonneau2008application, ni2020systematic}. Untrustworthy ML model may result in financial losses.

\subsection{Threat model}
\label{subsec:threat-model}

We assume a threat model for the ML marketplace in which model owners do not trust other entities with the confidentiality of their ML models. These ML models constitute their intellectual property, and thus they want to retain control over the execution of inference requests from clients. Model owners achieve this by running the models exclusively on their own cloud nodes. On the other hand, clients do not fully trust any single model owner to provide a correct inference result.

The adversary's goal is therefore to provide incorrect or malicious results to clients' inference requests whilst avoiding detection. The adversary could have corrupted one or more models at training time, \eg by poisoning or inserting backdoors or trojans~\cite{biggio2012poisoning, liu2017trojaning, shafahi2018poison}. Alternatively, the adversary may be a malicious or compromised model owner, \eg a disgruntled employee, or an external entity that has penetrated the defenses of the ML marketplace and can compromise a subset of the cloud nodes. Irrespective of the method of compromise, we assume the adversary can control and coordinate the behavior of all compromised models and nodes.

Nodes are distributed across multiple cloud accounts, \ie model owners have their own accounts with a cloud vendor. Each node has its own identity, \eg a public/private cryptography key pair, and the adversary is unable to obtain an honest node's private key. Nodes can establish secure communication channels, which cannot be compromised if both nodes are honest. We also assume that clients and model owners have verifiable identities~\cite{perlman1999overview, httpsope2:online}, which can be used to authenticate them and create secure channels.

%% file: 3-overview.tex
\section{Trustworthy ML Model Inference}
\label{sec:trustworthy_inference}

We discuss how to establish trustworthiness in an ML marketplace~(\S\ref{subsec:concpt-design}) and formalize the definition of trustworthy Byzantine agreement~(\S\ref{subsec:formal-definition}).

\subsection{Establishing trustworthiness}
\label{subsec:concpt-design}

Intuitively, a trustworthy inference result is a result that a client has obtained by correctly executing an inference request against an ML model. Clients can obtain trustworthy inference results in two ways: (i)~a result correctly executed by a trusted node; or (ii)~a collection of results from multiple nodes, with the client trusting the collective \emph{agreement} across the results.

\mypar{Trusted node execution} For a node to be trusted by the client, it must demonstrate its trustworthiness. This can be done because the node is controlled by the client or by another entity that is trusted by the client. In an ML marketplace, however, nodes are controlled by potentially untrusted model owners.   

Alternatively, inference computation on the node may be protected by a hardware security mechanism, such as  \emph{trusted execution environment}~(TEE)~\cite{volos2018graviton, costan2016intel}. A TEE leverages a root of trust from the hardware to shield the execution of sensitive computation and its data from the rest of the node, including higher privileged software layers. While TEEs have been proposed to make ML computation trustworthy~\cite{grover2018privado, narra2019privacy, narra2021origami, lee2019occlumency}, past TEE implementations have been shown to exhibit exploitable security vulnerabilities~\cite{buhren2021one, murdock2020plundervolt, van2020lvi, nilsson2020survey}.

\mypar{Trusted model agreement} An inference result can be untrustworthy if the model was trained on poisoned data~\cite{koh2022stronger, mei2015using}, a malicious actor changed the parameters during model training~\cite{biggio2012poisoning, liu2017trojaning, shafahi2018poison}, or the node which returned the inference result was untrustworthy. A technique to counter an untrustworthy inference result is to aggregate results from multiple independent models, \eg Bootstrap aggregation (bagging)~\cite{breiman1996bagging} combines multiple versions of models and increases accuracy in the presence of untrustworthy inference results~\cite{jia2020intrinsic, biggio2011bagging, smutz2016tree}. With bagging an ensemble classifier combines all the inference results and returns the prediction with which the largest number of input inference results agree.

\mypar{Byzantine model agreement} An important design decision is how agreement among the nodes with independent models is obtained. If the agreement is established by the client, it puts an additional burden on the client, because it changes the interface to the ML marketplace: instead of sending an inference request to a single API endpoint provided by the ML marketplace and receiving a response, the client must communicate individually with multiple nodes, aggregating the results from all their models. Therefore, it is preferable for model agreement to be provided by the ML marketplace itself.

Establishing agreement between models could be carried out by the ML marketplace provider, but this would make it a trusted entity. Instead, we explore a design that decentralises agreement among the nodes controlled by the model owners, without introducing any new trusted entities. In particular, the client does not have to trust the ML marketplace implementation to carry out model agreement correctly.

A Byzantine consensus protocol, \eg \emph{Practical Byzantine Fault Tolerance}~(PBFT)~\cite{pbft}, can obtain agreement on an execution result from a set of nodes, even if some nodes are malicious. A trustworthy ML marketplace could thus rely on the agreement of a set of nodes using Byzantine consensus. No individual untrustworthy model owner would then be able to convince the client to accept an incorrect inference result.

\subsection{Trustworthy inference through agreement}
\label{subsec:formal-definition}

We formalize the problem of providing a trustworthy ML marketplace as follows. We assume that a client submits an inference request against a set of models~$m_1 \ldots m_k \in \mathcal{G}$ that form a \emph{model group}~$\mathcal{G}$. A model group is a set of models that are semantically equivalent and can handle the same inference requests.

An inference request~$r$ is then served by an ML marketplace with nodes~$\mathcal{N}$. For a given model~$m \in \mathcal{G}$ and a node~$i \in \mathcal{N}$, let $q_i^m: \mathbb{R}^{u_m} \mapsto \mathbb{R}^{v_m}$ denote the implementation of the inference computation at node~$i$ for model~$m$, where input and output tensors are encoded into flattened arrays. (We require the input and output dimensions $u_m$ and $v_m$ to be the same at every node.)

The client obtains a set of inference results from $N=|\mathcal{N}|$~nodes, of which at most $f$~nodes may return an untrustworthy result. Therefore it is possible to construct a trustworthy inference result as long as there are at least $f$$+$$1$ results that the $f$$+$$1$~nodes have agreed on, because this quorum of nodes includes at least one trustworthy node.

To define which inference results are in agreement with each other, we introduce two parameters: (i)~a model-specific metric~$\delta_m: \mathbb{R}^{v_m} \mapsto \mathbb{R}$ that represents a meaningful distance between two results; and (ii)~a similarity threshold~$\epsilon_m$ that determines which results are sufficiently close to be considered in agreement. Both are provided by the ML marketplace as part of $\mathcal{G}$; a client may decide to select a different similarity threshold~$\epsilon_m$ to limit the maximum influence of incorrect or malicious nodes according to their own preferences. 

As a default, the model-specific metric~$\delta_m$ can be defined to be Euclidean distance
$ \delta_m(x,y)=\sqrt{\sum_{i=1}^{v_m} (x_i-y_i)^2 } $.
For simple models, such as classifiers that return a vector of probabilities for all possible classes, this measure suffices to exclude misclassifications; for more complex models, such as ones that return higher dimensional results, it is more appropriate to use a different distance measure such as Hausdorff~\cite{huttenlocher1993comparing} or complex wavelet similarity~\cite{wang2005translation}, as these are more robust to semantically irrelevant differences.

The distance metric~$\delta_m$, together with the threshold~$\epsilon_m$, can now be used to define the inference results that are in agreement. We denote $2^\mathcal{N}$ the powerset of $\mathcal{N}$, \ie the set of all subsets of $\mathcal{N}$, and $[2^\mathcal{N}]_f = \{C \in 2^\mathcal{N} \;|\; |C| \geq N-f\}$ its restriction to subsets with at most $f$ nodes removed.
Given a request~$r \in \mathbb{R}^{u_m}$ and subset $\mathcal{X} \in 2^\mathcal{N}$, we also define the diameter $\Delta_m(r,\mathcal{X})$ of the inference result set of $\mathcal{X}$ as:
\[ \Delta_m(r, \mathcal{X}) = \max_{i,j \in \mathcal{X}} \delta_m\left(q_i^m(r),q_j^m(r)\right) \]

The set of results from trustworthy nodes within the threshold $\epsilon_m$ is:
\[ Q_m(r) = \argmax_{\mathcal{X} \in [2^\mathcal{N}]_f} \{ |\mathcal{X}| \; | \; \Delta_m(r,\mathcal{X}) \leq \epsilon_m \} \]
This definition selects the largest subset of at least $N-f$ nodes whose diameter is within the threshold, rather than the one with minimal diameter overall (which may unnecessarily exclude legitimate results). While $f+1$ nodes may be enough for agreement, it would make the definition more brittle, as there could be two sets of equal size (greater than $f+1$) of diameter smaller than $\epsilon_m$ if all malicious nodes skew their result towards a honest outlier. Clients can set a small $\epsilon_m$ to increase consistency between results. However, $Q_m(r)$ may fail to yield a trustworthy subset if honest nodes are too inconsistent and the client may consider whether a larger $\epsilon_m$ is acceptable. 

Note that $Q_m(r)$ only contains the set of nodes that produce a trustworthy result with respect to $\delta_m$ and $\epsilon_m$. In addition, clients may also expect the ML marketplace to return a single aggregate inference result based on $\{ q_i^m(r), i \in Q_m(r)\}$, \eg using an average or computed $\delta_m$-center approximations.

\section{\sys API}
\label{sec:api}

Next, we describe the API for our trustworthy ML marketplace called \sys. The API in \T\ref{tab:api_table} shows the interface through which \emph{model owners} manage models and \emph{clients} execute trustworthy inference requests.

\begin{table}[tb]
    {\scriptsize
      \begin{tabular}{lp{3.8cm}}
      \toprule
        \textbf{API function}  & \textbf{Description} \\
        \midrule
        \texttt{\textbf{get\_API\_endpoints}()} & \multirow{2}{4.25cm}{Obtains API endpoint and node keys.} \\
        \hspace{2ex}$\rightarrow$ \texttt{([API\_URL], [pub\_key])} & \\

        \texttt{\textbf{get\_model\_groups}() $\rightarrow$ ([mdl\_grp])} & Obtains model groups. \\

        \midrule

      \texttt{\textbf{upload\_model}(model\_URL, [param])} & Uploads model to marketplace. \\
      \texttt{\textbf{retire\_model}(model\_URL)} & Marks model as ready to retire. \\

        \midrule
        
        \texttt{\textbf{define\_model\_group}(mdl\_grp, } & \multirow{2}{4.25cm}{Defines \texttt{models} in \texttt{model\_group} with distance function.}\\
        \hspace{2ex}\texttt{[model\_URL], dist\_fn)}  & \\

      \texttt{\textbf{activate\_model\_group}(mdl\_grp)} & Activates model group. \\

      \texttt{\textbf{retire\_model\_group}(mdl\_grp)} & Retires model group. \\ 
        
        \midrule
        
        \texttt{\textbf{request\_inf}(mdl\_grp, input, $\epsilon$)} & \multirow{2}{4.25cm}{Executes request against model group; return results and certificate.} \\
        \hspace{2ex}$\rightarrow$ (\texttt{[inf\_res]}, \texttt{inf\_cert}, dist\_fn) & \\

        \texttt{\textbf{verify\_cert}(inf\_req, [inf\_res],} & \multirow{2}{4.25cm}{Verifies validity of inference certificate.} \\
          \hspace{2ex}\texttt{inf\_cert, [pub\_key])} $\rightarrow$ \texttt{bool} &  \\

        \bottomrule
      \end{tabular}
    }
  \caption{{\System} API}
  \label{tab:api_table}
\end{table}

Both model owners and clients must first obtain an API endpoint. The call \texttt{get\_API\_endpoints} contacts a trusted service at a well-known location (\eg DNS) and returns a list of URL endpoints (\texttt{API\_URL}) with their public keys. These identify $N$~nodes of the \sys service, of which up to $f$ may not be trustworthy. API requests can be sent to any endpoint URL.

\myparr{Model owners} maintain cloud nodes, which execute inference requests against their models.  They add or update models using the \texttt{upload\_model} call, which includes a URL of a model in the ONNX format~\cite{onnx} and further meta-data parameters  to assign the model to model groups.

Before clients can issue inference requests against a new model, the model must be added to a model group. The models included in a model group are selected by the ML marketplace based on the parameters supplied by the model owners. The \texttt{define\_model\_group} call specifies the models in the group and its distance function, which takes a set of inference results and returns those it deemed to be trustworthy (see~$Q_m(r)$  in \S\ref{subsec:formal-definition}). The group is activated using \texttt{activate\_model\_group}.

\Cref{fig:api-code} shows an example how a model owner uploads a model, which is then added to a model group called \texttt{ResNet}. The model group uses a distance function that returns the largest set of results where all results are within a threshold of $0.2$ (line~\ref{fig:api-code:dist}). The model owner uploads the model (line~\ref{fig:api-code:load}), and the ML marketplace defines and activates the model group~(lines~\ref{fig:api-code:define}--\ref{fig:api-code:activate}). The distance function assumes a classification model that returns a confidence score for each label.

\begin{listing}[t] 
  \begin{lstlisting}[linewidth=\columnwidth, basicstyle=\scriptsize\ttfamily\mdseries, numberstyle=\tiny, numbersep=-1em, tabsize=2, aboveskip=0pt, belowskip=0pt, breaklines=true, frame=lines]
  [API_URL], [pub_key] = get_API_endpoint()
  model_URL = "https://[@($...$)@]/resnet101.onnx"
  upload_model(model_URL, [p0, p1])@(\label{fig:api-code:load})@
\end{lstlisting}
\begin{lstlisting}[linewidth=\columnwidth, basicstyle=\scriptsize\ttfamily\mdseries, numberstyle=\tiny, numbersep=-1em, tabsize=2, aboveskip=0pt, belowskip=0pt, breaklines=true, firstnumber=4]
  dist_fn = lambda results,dist=0.2:max(filter(lambda set:max(set)-min(set)<dist,powerset(results)), key=len)@(\label{fig:api-code:dist})@
  define_model_group("ResNet",[model_URL,@($...$)@],dist_fn)@(\label{fig:api-code:define})@
  activate_model_group("ResNet")@(\label{fig:api-code:activate})@
\end{lstlisting}\vspace{3pt}
\begin{lstlisting}[linewidth=\columnwidth, basicstyle=\scriptsize\ttfamily\mdseries, numberstyle=\tiny, numbersep=-1em, tabsize=2, aboveskip=0pt, belowskip=0pt, breaklines=true, frame=lines, firstnumber=7]
  [model_group] = get_model_groups()@(\label{fig:api-code:get_model_groups})@
  [inf_res], inf_cert, _ = request_inf(model_group, data)@(\label{fig:api-code:request_inf})@
  verify_cert(inf_req, [inf_res], inf_cert, [pub_key])@(\label{fig:api-code:verify_cert})@\end{lstlisting}
  \caption{Example use of \sys API}\label{fig:api-code}
\end{listing}

\myparr{Clients} issue inference requests against a model group using the \texttt{request\_inf} call (line~\ref{fig:api-code:request_inf}), which takes a similarity threshold~$\epsilon$ to the distance function (see~$\epsilon_m$ in \S\ref{subsec:formal-definition}); if $\epsilon$ is left blank, the default is used. After executing the inference request, \sys returns multiple inference results (\texttt{[inf\_res]}) from the nodes that execute the request and an \emph{inference certificate}~\texttt{inf\_cert}. The inference certificate provides a proof that the nodes agreed on the outcome of the inference execution.

Client verifies the certificate together with the results by calling \texttt{verify\_cert}~(line~\ref{fig:api-code:verify_cert}). It checks the signatures of nodes over the inference results and their signatures over the results from other nodes (see~\S\ref{sec:design:attest}). A node's signature over another node's result shows their belief that the node's result is trustworthy, which we refer to as an \emph{attestation}. The function first checks that at least $N$$-$$f$~nodes returned inference results, and then validates that each inference result is signed by the node that produced it. For every signed result, the function checks that at least $f$$+$$1$~attestations with valid signatures exist. This ensures that every result is attested by at least one trustworthy node, \ie more than $f$~nodes attested the results.


%% file: 5-design.tex
\section{\sys Design}

We give an overview of \sys{}'s design~(\S\ref{sec:design:overview}) and explain how it provides trustworthy inference results using an execute~(\S\ref{sec:design:execute}), agree~(\S\ref{sec:design:agree}), and attest~(\S\ref{sec:design:attest}) strategy. 

\subsection{Overview}
\label{sec:design:overview}

\begin{figure}[tb]
  \centering
  \includegraphics[width=\columnwidth]{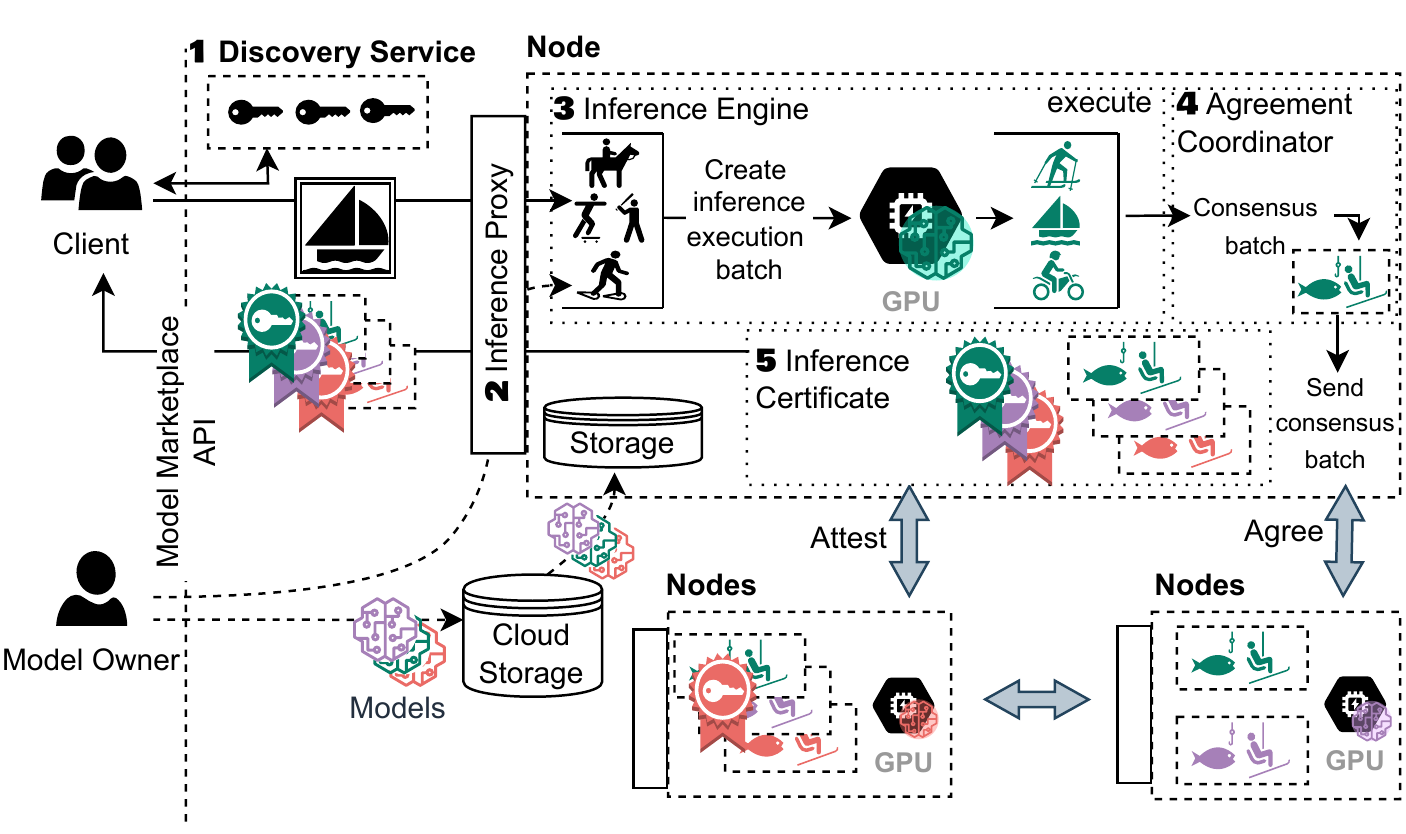}
  \caption{\sys design}\label{fig:overview}
\end{figure}

As shown in \F\ref{fig:overview}, a \sys deployment consists of $N$~nodes, which are controlled by distinct model owners. To provide a trustworthy ML marketplace~(see~\S\ref{subsec:formal-definition}), each node manages models from model groups and \emph{executes} inference requests against them. The results provided by different nodes must then \emph{agree} with each other. Agreement can only be achieved if inference requests are \emph{ordered} consistently with respect to model updates, otherwise nodes may execute inference requests against different versions of the same model. Finally, the agreement must be \emph{attested} to prove it to the client.

The design of \sys realizes the above functionality by following an \emph{execute/agree/attest} strategy, which separates (i)~the execution of inference requests and model updates from (ii)~the agreement and ordering of the results; and (iii)~their attestation for the client. By separating these phases, it becomes possible to achieve higher request throughput, because operations can be batched independently: first requests are grouped into \emph{execution batches}, which are executed in parallel on all $N$~nodes; and then results are collected and grouped into \emph{agreement batches}, which agree on a trustworthy result using a Byzantine consensus protocol.

\sys selects the maximum execution batch size based on the level of GPU parallelism; and the maximum agreement batch size based on the network latencies between nodes. We experimentally demonstrate the performance benefit of decoupling execution from agreement in \S\ref{subsec:impact-batching}.

Each node is identified uniquely by its public/private signing keys. Clients use a trusted \textbf{discovery service}~\myc{1} to learn about node identities. They can send API requests to any node. A node receives requests through its \textbf{\infproxy}~\myc{2}, which acts as an endpoint for the \sys API~(see~\S\ref{sec:api}). A node uses its public key to establish a secure TLS communication channel with the client and authenticates it. 

The \infproxy forwards requests to the \textbf{inference engines}~\myc{3} of other nodes. The inference engine batches requests and uses a GPU inference library to \emph{execute} batches on GPUs. After execution, the \textbf{agreement coordinator}~\myc{4} \emph{agrees} on the order of inference requests and model group updates, and on their execution results through Byzantine consensus. Finally, the \infproxy \emph{attests} the result by constructing an \textbf{inference certificate}~\myc{5}. The inference certificate includes appropriate attestations that prove the result's trustworthiness.

\subsection{Execute}
\label{sec:design:execute}

After receiving an API call (see~\Cref{tab:api_table}), the \infproxy forwards it to \emph{all} other nodes. Requests for defining model groups (\texttt{define\_model\_group}) and executing inference requests~(\texttt{request\_inf}) are then executed by the inference engine on each node, as described below; requests for activating/retiring model groups~(\texttt{activate}/\texttt{retire\_model\_group}) are directly ordered by the agreement coordinator (see~\S\ref{sec:design:agree}).

\mypar{Defining model groups} On a \texttt{define\_model\_group} request, an inference engine loads the models from the model group into GPU memory, which makes them available for inference execution. The models are retrieved from the specified URLs and cached locally by the node.

A group of nodes controlled by the same model owner~$d$ loads the models from the model group. If the number of models, $|\mathcal{G}|$, is larger than $n$, the nodes load disjoint partitions of models; otherwise, some models are replicated across nodes. The inference engine selects deterministically which $c = \lceil d / |\mathcal{G}| \rceil$ models to load for each group of nodes. The ordered list of models in the model group is divided into $c$-sized groups, and inference engines are assigned groups based on a total order imposed by the nodes' public keys.

\mypar{Executing inference requests} The inference engine executes requests in batches by grouping them into \emph{execution batches} that refer to the same model. There is a trade-off when setting the maximum batch size: while larger batches help exploit GPU parallelism, they increase latency, consume more GPU memory and may become bottlenecked by PCIe bandwidth. The actual used batch size also depends on the number of concurrently submitted inference requests that refer to the same model. We explore the impact of the maximum execution batch size on performance in \S\ref{subsec:impact-batching}.

\sys guarantees that inference requests are always executed against the \emph{latest} activated model version. This model, however, is only determined at agreement time after requests have been ordered (see~\S\ref{sec:design:agree})---at execution time, multiple model versions belonging to the same model group may exist. To overcome this problem, the inference engine executes such requests against \emph{all} versions. This ensures that, even if a new model group version is activated before the request is ordered, the request will have been executed against that version.

\subsection{Agree}
\label{sec:design:agree}

Parallel batches execute on all nodes, and nodes must agree on the order of inference requests and model group activations/retirements to return consistent and trustworthy results to clients. \sys uses a Byzantine consensus protocol~(PBFT~\cite{pbft}) to order requests, which mitigates $f$~untrustworthy nodes that return incorrect results. The inference results produced by executing the ordered requests establish trusted agreement (see~\S\ref{subsec:concpt-design}), which distributes trust among nodes.

\sys uses the PBFT protocol to choose a \emph{primary} node, which is determined based on a monotonically increasing counter called view~$v$. An \emph{agreement coordinator} on the primary node orders inference requests relative to model group updates. It creates \emph{agreement batches} and assigns a monotonically increasing sequence number~$n$ to each batch. Each agreement batch contains an ordered list of inference requests and model updates.

\mypar{Ordering requests} The agreement coordinator on the primary node proposes an agreement batch, which contains an ordered list of requests~$\mathcal{O}$. It includes the agreement batch in a \msg{PRE-PREPARE} message with the following format: $\langle \msg{PRE-PREPARE}, v, n, H(\mathcal{O}), \mathcal{R}_r \rangle_{\sigma}$. The message specifies which node's agreement coordinator created the ordering by including the view~$v$ and records the agreement batch's sequence number~$n$ and its hash~$H(\mathcal{O})$. It also includes $\mathcal{R}_r$, which is the root of a Merkle tree~$\mathcal{R}$ that is constructed by the agreement coordinator over the inference requests and results. As explained in \S\ref{sec:design:attest}, the signed Merkle tree root is used as an optimization to reduce the number of signatures that must be verified. Finally, $\sigma$ is a signature over the whole message. The \msg{PRE-PREPARE} message, $\mathcal{R}$, and $\mathcal{O}$ are sent to the agreement coordinators on all other nodes, which act as \emph{backup nodes}.

When a backup node's agreement coordinator receives the primary's agreement batch, it constructs its own agreement batch and sends it in a \msg{PREPARE} message to all nodes. This enables the nodes to obtain the required number of results to check if they are trustworthy. The \msg{PREPARE} message has the following format: $\langle \msg{PREPARE}, v, n, j, i, \mathcal{R}_r \rangle_{\sigma_i}$. It includes $\mathcal{R}_r$, which is the root of a Merkle tree whose leaves are the batch's inference requests and results. The \msg{PREPARE} also includes ordering information, but the agreement coordinator removes the need to include $\mathcal{O}$ by including $j$, a hash of the \msg{PRE-PREPARE} message. Finally, the message contains $i$, the public key of the node that created the message, and is signed by public key~$\sigma_i$. Again, the agreement coordinator sends the \msg{PREPARE} message and $\mathcal{R}$ to all agreement coordinators.

When ordering agreement batches, the primary's agreement coordinator considers each request in the batch and updates its state: for an \texttt{activate\_model\_group} request, it updates the active version of the model group; \texttt{retire\_model\_group} deactivates all versions of the model group, reclaiming used GPU memory; for a \texttt{req\_inference} request, the coordinator considers all inference results previously produced for different versions of the model group. Since the inference request has now been ordered with respect to previous \texttt{activate\_model\_group} requests, the coordinator only retains the result from the model group's latest version.

The optimal agreement batch size depends on network latency between the nodes. A larger batch size reduces the impact of the network latency, but it increases the request latency that clients experience. The agreement batch size grows linearly with the result size, as they are included in the batch. Inference results can be large, \eg a single ResNet model may return a 4000\unit{byte} inference result. Our experimental results show that, if the agreement batch size is chosen correctly, network latency does not limited overall throughput (see~\S\ref{subsec:scalability}).

\mypar{Achieving agreement} When a backup node's agreement coordinator receives the primary's agreement batch, it applies the ordering information: for each request in the batch, it selects the appropriate inference result or updates its state. As it sends its inference results in an agreement batch to the other agreement coordinators, each coordinator receives the $N$$-$$f$~inference results needed to produce a trustworthy inference result~(see~\S\ref{subsec:formal-definition}).

To produce a trustworthy inference result, the agreement coordinator applies the \emph{distance function} after obtaining $N$$-$$f$~inference results.
From the $N$$-$$f$ inference results, the function removes ones that are not subject to the agreement bounds. For each request in the agreement batch, the agreement coordinator applies the distance function of the active version of the model group at the time when the request was ordered. Note that the distance function is executed within a sandbox, isolating it from the node's execution and bounding its resources.

\subsection{Attest}
\label{sec:design:attest}

After receiving a set of trustworthy inference results, the \infproxy must prove the results' trustworthiness to the client, which requires that at least one trustworthy node has attested the results. \sys thus provides the client with $f$$+$$1$~attestations of which at least one is trustworthy.

An attestation is created by the agreement coordinator, which signs the inference results that it obtained after applying the distance function. The results are signed by the node's private key and are broadcast to all agreement coordinators. Based on this, an inference proxy can construct an inference certificate, which shows that at least $f$$+$$1$~nodes have agreed and attested at least $N$$-$$f$~results.

\mypar{Attesting results} For an agreement coordinator to create an attestation, it must obtain $N$$-$$f$~inference results. To prove to the client that the results were produced on multiple nodes, they must be signed by their private keys. Instead of signing all inference results in the agreement batch separately, the agreement coordinator constructs a Merkle tree~$\mathcal{R}$ in which each leaf is an inference request and result. This allows the agreement coordinator to sign the Merkle tree root~$\mathcal{R}_r$ once. By including the Merkle authentication path in the inference certificate, the signature is linked to the inference result.

A node's agreement coordinator waits until it has received $N$$-$$f$~agreement batches before attestation~(see~\S\ref{sec:design:agree}). It then signs a hash of the results within the batch and sends it in a \msg{COMMIT} message to all agreement coordinators. The \msg{COMMIT} message has the format: $\langle \msg{COMMIT}, v, n, j, i, \mathcal{A}_r \rangle_{\sigma_i}$. As before, including the root of a Merkle tree~$\mathcal{A}$ avoids the need to sign each attestation, because the leaves are the hashes of the inference requests and results being attested. If the agreement coordinator attests all results in an agreement batch, it includes $\mathcal{R}_r$ from the agreement batch in $\mathcal{A}$ instead of adding individual results. The agreement coordinator sends the \msg{COMMIT} message and $\mathcal{A}$ to all agreement coordinators.

If an agreement coordinator receives an attestation for an inference result that it has not obtained, it requests that the sender forward the missing result and the \msg{PREPARE} message. It waits for a response before it considers the attestation as completed. Requests are ordered after the agreement coordinator has received $N$$-$$f$$-$$1$ \msg{COMMIT} messages.

\mypar{Creating inference certificates} After obtaining $N$$-$$f$~attestations, an \infproxy creates an \emph{inference certificate} with the following format: $\langle v, n, H(\mathcal{O}), \mathcal{S}, \mathcal{P}, \mathcal{D} \rangle$. It contains $\mathcal{S}$, the signatures over the \msg{PRE-PREPARE}, \msg{PREPARE}, and \msg{COMMIT} messages; $\mathcal{P}$, the authentication paths from the Merkle tree~$R$; and $\mathcal{D}$, the authentication paths over the attestations~$\mathcal{A}$. After creating the certificate, the \infproxy forwards it, together with the inference results, to the client.

\input{algorithms/verify-cert.tex}

As described in \S\ref{sec:api}, the client can now verify the inference certificate and results using the \texttt{verify\_cert} function. \Cref{alg:verify-cert} provides the functions' pseudocode: the client checks that the nodes signed their own inference results, confirms that at least $N$$-$$f$~results were returned~(line~\ref{veri-cert-res-count}), and iterates through the inference results (line~\ref{veri-cert-iter-results}); it obtains the Merkle tree root~$\mathcal{R}_r$ using the authentication path provided in $\mathcal{P}$~(line~\ref{veri-cert-mr-res}), and uses this to verify the node's signature over its result~(line~\ref{veri-cert-check-sig}).

The client then confirms that the results were attested by at least one trustworthy node. It checks that a minimum node number have attested each result~(line~\ref{veri-cert-endorse-count}). For each attestation~(line~\ref{veri-cert-iter-endorse}), it reconstructs the root of each node's Merkle tree~(line~\ref{veri-cert-mr-endorse}) and verifies the signatures~(line~\ref{veri-cert-verify-endorse}).

\subsection{Discussion}
\label{sec:design:discussion}

\mypar{Security analysis} We discuss how \sys guarantees the trustworthiness of inference results.

\mypari{Untrustworthy \infproxy} When a malicious \infproxy receives a client request, it either: (i)~forwards the request to a subset of nodes, which means that trustworthy nodes can forward the request to one another~\cite{castro2001practical, clement2009making}; or (ii)~discards the request, which means that the client will not receive a valid inference certificate. After a timeout, the client re-sends the request to another \infproxy. After at most $f$$+$$1$ tries, the client will find a trustworthy \infproxy. 

\myparii{Untrustworthy nodes}, controlled by a misbehaving model owner, may tamper with the execution and ordering of requests. Since the PBFT protocol orders requests, the produced results are from consistently versioned models in a model group. If the primary node misbehaves, it is re-assigned through PBFT's view change protocol. The primary is determined based on the monotonically-increasing view~$v$, which increments each time the primary node is changed. To decide on the primary node, the nodes' public keys are ordered lexicographically, and the $p^{\mathrm{th}}$ node from the list is chosen as the primary where $p\;=\;v\;\mathrm{mod}\;N$. A view change is triggered by backup nodes if they do not receive $N$$-$$f$ attestations within a timeout after receiving the request.

\mypari{Attacking accuracy} Untrustworthy nodes may try to undermine result accuracy of the model group~(see~\S\ref{subsec:impact-failure}). For example, a malicious node may slightly perturb the generated results to reduce model group accuracy~\cite{yuan2019adversarial}. The distance function mitigates this issue by bounding the tolerated divergence between models.

\mypar{Result aggregation} By design, \sys does not aggregate the model results in the model group. It leaves this task to the client, because the \infproxy cannot be trusted to aggregate results correctly. \sys could be extended to provide trustworthy aggregated results by obtaining an aggregation function $\Sigma_m$ from the ML marketplace. It would then add an extra consensus round in its agree phase to ensure that $f+1$~nodes agree over the value of $\Sigma_m(Q_m(r))$.

\mypar{Non-determinism} We assume that the execution of an inference request on the same GPU with the same model returns the same result. However, if the same inference request is executed on different hardware, results may differ. A model owner who is aware that inference computation executes on heterogenous hardware may provide a distance function that accounts for this.

%% file: algorithms/verify-cert.tex
\SetKwFunction{FVerifyCert}{\textsf{verify\_cert}}
\SetKwFunction{FGetMerkleRoot}{\textsf{\footnotesize get\_merkle\_root}}
\SetKwFunction{FHashPrePrepare}{\textsf{\footnotesize hash\_PrePrepare}}
\SetKwFunction{FVerifySig}{\textsf{\footnotesize verify\_sig}}
\SetKwFunction{FGetAttestations}{\textsf{\footnotesize get\_attestations}}

\begin{algorithm}[tb]
  \footnotesize
  \caption{\textbf{Verify inference certificate} (It uses the public
    keys~$\mathcal{K}$ from the discovery service.)}\label{alg:verify-cert}
  \SetAlgoNoEnd
  \DontPrintSemicolon 
  \SetNoFillComment
  \SetInd{0.4em}{0.4em}
\BlankLine
  \KwOn({\FVerifyCert{$\langle v, n, \mathsf{H}(\mathcal{O}), \mathcal{S}, \mathcal{P}, \mathcal{D} \rangle, \mathcal{R}, r, \mathcal{K}, f$}})
  { 
    \scriptsize\textsf{ 
      \lIf{$|\mathcal{R}| < N-f$}
      {
        \KwRet{$\mathsf{False}$}
      }
      \label{veri-cert-res-count}
      $\mathsf{h_{pp}} \gets \mathsf{H}(\langle $PRE-PREPARE$ \; v, n, H(\mathcal{O}), \mathcal{R}_{i_r} \rangle_{\mathcal{S}_i}$) \;
      \ForEach{$i, \mathcal{R}_i \in \mathcal{R}$}
      {
        \label{veri-cert-iter-results}
        $M_r \gets $\FGetMerkleRoot{$i, \mathcal{P}, r, \mathcal{R}_i$} \;
        \label{veri-cert-mr-res}
        \lIf{$\neg$ \FVerifySig{$v, n, H(\mathcal{O}), M_r, i, \mathsf{h_{pp}}, \mathcal{S}_i, \mathcal{K}_i$}}
        {
          \KwRet{$\mathsf{False}$}
        }
        \label{veri-cert-check-sig}
        $\mathsf{attestations_{\mathcal{R}_i}} \gets $ \FGetAttestations($\mathcal{D}, \mathcal{R}_i$) \;
        \lIf{$|\mathsf{attestations_{\mathcal{R}_i}}| \leq f$}
        {
          \KwRet{$\mathsf{False}$}
        }
        \label{veri-cert-endorse-count}
        \ForEach{$\langle i, M_{path} \rangle \in \mathsf{attestsations_{\mathcal{R}_i}}$}
        {
          \label{veri-cert-iter-endorse}
          $M_r \gets $\FGetMerkleRoot{$M_{path}, r, \mathsf{\mathcal{R}_i}$} \;
          \label{veri-cert-mr-endorse}
          \lIf{$\neg$ \FVerifySig{$v, n, M_r, i, \mathsf{h_{pp}}, \mathcal{S}_i, \mathcal{K}_i$}}
          {
            \KwRet{$\mathsf{False}$}
          }
          \label{veri-cert-verify-endorse}
        }
      }
      \KwRet{$\mathsf{True}$}
    }
  }
\end{algorithm} 


%% file: 6-evaluation.tex
\section{Evaluation}

We evaluate \sys to explore the performance overhead of providing trustworthy ML inference results (\S\ref{subsec:comparison}), the benefit of separating execution and agreement batching (\S\ref{subsec:impact-batching}), the effect of model agreement (\S\ref{subsec:agg-vs-rep}) and model updates (\S\ref{subsec:model-update}), and \sys{}'s scalability (\S\ref{subsec:scalability}). We finish by considering the impact of untrustworthy nodes on accuracy (\S\ref{subsec:impact-failure}) and of inference request and result sizes (\S\ref{subsec:impact-req-res-size}).

\subsection{Experimental setup}
\label{sec:eval:setup}

We implement \sys in approximately 5,000 lines of C++ code using EverCrypt's SHA functions~\cite{protzenko2020evercrypt}, the MerkleCPP library~\cite{microsof86:online}, the MbedTLS library for client connections~\cite{embedtls}, and secp256k1 for all signatures~\cite{wuille2018libsecp256k1}. It also uses the ONNX Runtime (v1.8.1)~\cite{onnx_runtime} as part of its inference engine to execute inference requests on GPUs.

\mypar{Testbeds} Our setup consists of three environments: (a)~\textsf{cluster}: a dedicated 5-machine cluster, each with an 8-core 3.7-Ghz Intel E-2288G CPU with 16\unit{GB} of RAM and a 40\unit{Gbps} network with full bi-section bandwidth; (b)~\textsf{cloud/single-site}: a single datacenter cluster in the Azure cloud (East US) with NC12s\_v3 VMs, each with a 12-core 2.60-Ghz Intel E5-2690 CPU with 224\unit{GB} of RAM and 2 NVIDIA V100 GPUs with 16\unit{GB} of RAM; and (c)~\textsf{cloud/multi-site}---a WAN configuration using the same VMs across 3 Azure regions (East US, East US 2, US South Central). All machines run Ubuntu Linux~18.04.6 LTS.

We set the execution batch size to 4, as larger batch sizes do not improve performance with our GPUs. In the single-site environments, we use an agreement batch size of 25 with a batch pipeline of 2; in \textsf{cloud/multi-site}, we use a batch size of 50 with a pipeline of 4.

\mypar{Workloads} To emulate a distribution of DNN models in a cloud-based ML marketplace, we use 42 ImageNet models taken from the ONNX and PyTorch model zoos~\cite{GitHubon40:online, 10ModelZ20:online} (see~\Cref{tab:models}). These models vary in size from several to hundreds of MBs, covering various complexities and depths. We update all models to accept batches of inference requests. We employ a set of 1000~images taken from the ImageNet validation dataset~\cite{russakovsky2015imagenet} as input to inference requests.

\begin{table}[tb]
  \centering
  \footnotesize%
  \begin{tabular}{lr|lr}\toprule
    \textbf{Model family} & \textbf{\#Vars} &  \textbf{Model family} & \textbf{\#Vars} \\
    \midrule
    ResNet~\cite{he2016deep, he2016identity}       & 17  & SqueezeNet~\cite{iandola2016squeezenet}   & 3                \\
    VGG~\cite{simonyan2014very}          & 10 & AlexNet~\cite{krizhevsky2012imagenet}      & 1                \\
    DenseNet~\cite{huang2017densely}     & 5  & GoogleNet~\cite{szegedy2015going}    & 1                \\
    Inception~\cite{szegedy2016rethinking}    & 1 & MnasNet~\cite{tan2019mnasnet}      & 2                \\
    ResNext~\cite{xie2017aggregated}      & 2     &&          \\
    \bottomrule
  \end{tabular}
  \caption{DNN models and variants used in evaluation}%
  \label{tab:models}
\end{table}

\subsection{Inference throughput and latency}
\label{subsec:comparison}

\begin{figure}[tb]
  \centering
  \includegraphics[width=1.0\columnwidth]{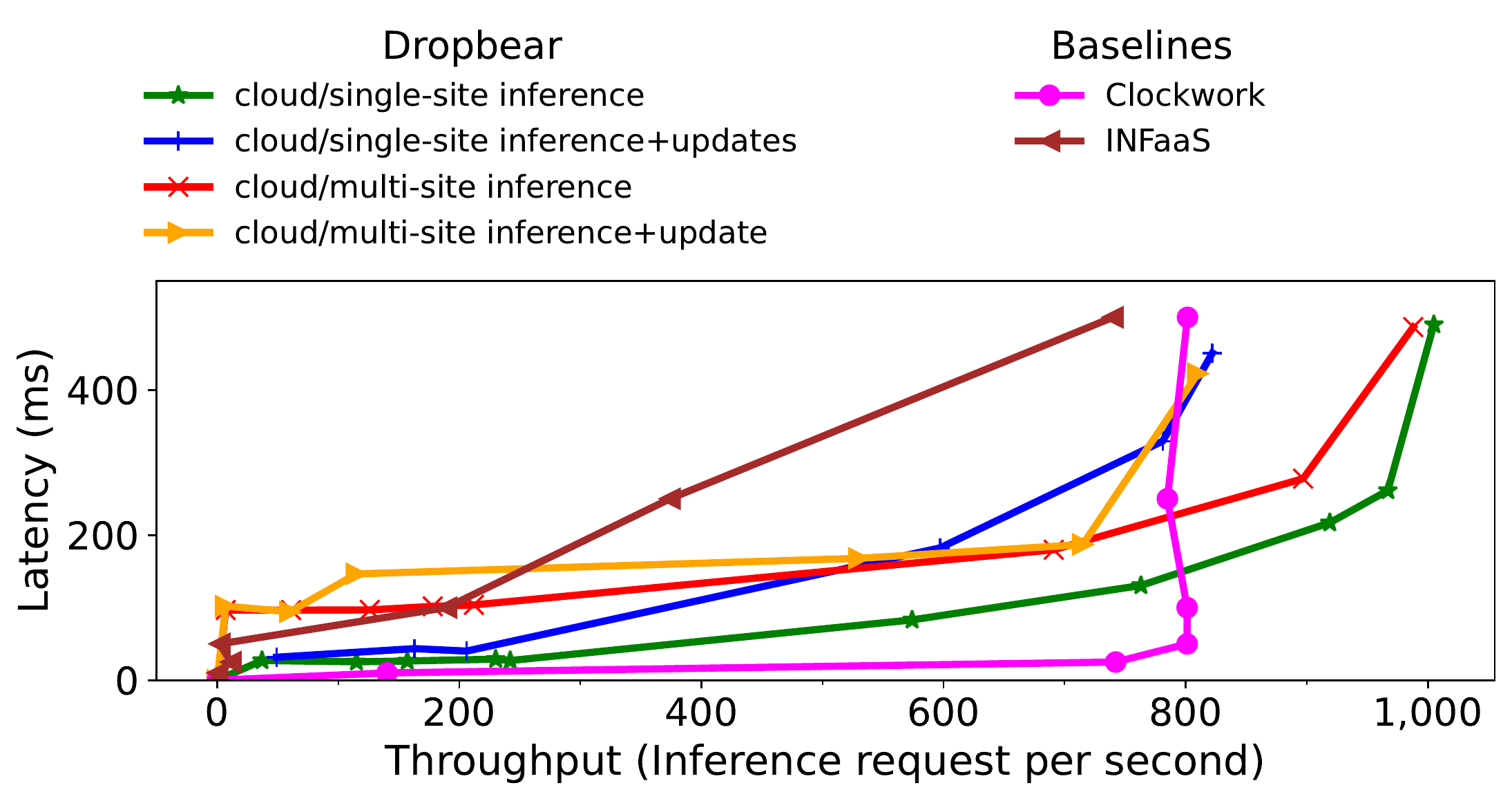}
  \caption{Request throughput versus latency \normalfont{(All systems use 15~ResNet50 instances. Clockwork and INFaaS use 1~worker and the reported latency as their target SLO. \sys is deployed with $f$$=$$1$.)}}
  \label{fig:comparison}
\end{figure}

We measure the impact on throughput and latency when providing trustworthy inference requests using \sys. We compare against two state-of-the-art distributed inference systems, which do not support trustworthy inference through agreement: (i)~Clockwork~\cite{gujarati2020serving} focuses on predictable inference latencies; and (ii)~INFaaS~\cite{romero2021infaas} selects models based on the SLOs of inference requests.

Similar to the authors of Clockwork, we were unable to deploy prior systems on our testbeds: Clockwork requires more GPU memory (32\unit{GB}) than is available to us; INFaaS only supports AWS. Since the hardware configurations from the respective papers (Clockwork: Dell PowerEdge R740 servers with 32 CPU cores, 32\unit{GB} RAM, and 2 NVIDIA V100 GPUs; InFaaS: AWS m5.24xlarge and p3.2xlarge VMs) are similar to ours, we reproduce their results for comparison. Note that all three systems use workers with 2 GPUs each. While the \sys deployment has 4$\times$ more workers, it must also execute each inference request against 4 models instead of 1. 

We consider two workloads for \sys: \textsf{inference} only consists of inference requests; \textsf{inference+updates} also includes updates to around 10\% of the model groups in parallel to executing inference requests. Clockwork and INFaaS do not perform any model updates.

\Cref{fig:comparison} shows a throughput vs.\ latency plot. \sys's peak throughput with inference requests only is 1005\unit{request/s} with a latency of 489\unit{ms} and 988\unit{request/s} with a latency of 487\unit{ms} in the \textsf{cloud/multi-site} configuration, respectively. With model updates, \ie ordering updates with inference requests, the throughput drops to 822\unit{request/s} with an average latency of 451\unit{ms} and 810\unit{request/s} with a latency of 422\unit{ms} for \textsf{cloud/multi-site}, respectively. \sys's throughput is bounded by the request execution on the GPU, based on the requests included in an execution batch and the consumed PCIe bandwidth when models are loaded (see~\S\ref{subsec:impact-batching}). The increase in latency arises from the cryptography operations to verify inference requests and produce inference certificates.

In comparison, Clockwork achieves a peak throughput of 801\unit{requests/s} with a latency of 50\unit{ms} without guaranteeing the trustworthiness of results. Clockwork is unable to raise throughput beyond this point, even when its SLO budget is increased 10$\times$ to 500\unit{ms}. INFaaS achieves a comparable throughput of 739\unit{requests/s} with a 500\unit{ms} latency.

Despite the extra communication and cryptographic overhead of the execute/agree/attest approach, \sys manages to achieve comparable throughput. This is due to the fact that the bottleneck is the expensive GPU inference computation, which hides the additional cost of trustworthy agreement. \sys's slightly higher throughput can be explained by the systems' different design goals: Clockwork makes a throughput trade-off to increase latency predictability, and does not use multiple threads to send inference requests to the same GPU; INFaaS trades-off throughput to obtain the highest inference accuracy within a time budget and uses the Triton Inference Server to execute requests.

To put \sys{}'s throughput into perspective, we use a micro-benchmark to measure the maximum throughput of inference requests that the ONNX Runtime~\cite{onnx_runtime} handles on one Azure NC12s\_v3 VM. With pure requests, \sys{}'s throughput reported above is within 4\% of this maximum. This demonstrates that the \sys{}'s performance is limited by the GPU computation performed by the GPU inference library.

\subsection{Impact of batching}
\label{subsec:impact-batching}

\begin{table}[tb]
\centering
  \footnotesize\addtolength{\tabcolsep}{-3pt}
  \begin{subtable}[t]{.49\linewidth}
    \centering
    \begin{tabular}{lc|ccccc} 
    \toprule  
    \multicolumn{7}{c}{\small\textbf{Execution batch size}}  \\
    && 1 & 2 & 4 & 8 & 16 \\
    \midrule
    \multirow{1}{*}{\notsotiny\rotatebox[origin=c]{90}{\textbf{\# Model groups}}} 
    & 10 & 321 & 592 & 877 & 888 &  -  \\
    & 20 & 350 & 583 & 817 & 785 &  -  \\
    & 30 & 280 & 533 & 721 &  -  &  -  \\
    & 42 & 289 & 471 & 633 &  -  &  -  \\
    \bottomrule
    \end{tabular}
    \caption{\textnormal{inference+updates}}
    \label{tab:impact:batching:inf}
  \end{subtable}
  \begin{subtable}[t]{.49\linewidth}
    \centering
    \begin{tabular}{lc|ccccc} 
    \toprule  
    \multicolumn{7}{c}{\small\textbf{Execution batch size}}  \\
    && 1 & 2 & 4 & 8 & 16 \\
    \midrule
    \multirow{1}{*}{\notsotiny\rotatebox[origin=c]{90}{\textbf{\# Model groups}}} 
     & 10 & 302 & 521 & 844 & 924 & 904 \\
     & 20 & 333 & 564 & 818 & 900 &  -  \\
     & 30 & 278 & 495 & 735 & 722 &  -  \\
     & 42 & 291 & 489 & 681 & 696 &  -  \\
    \bottomrule
    \end{tabular}
    \caption{\textnormal{inference}}
    \label{tab:impact:batching:inf-model-update}
  \end{subtable}
  \caption{Batch sizes and model group counts \normalfont{(``-'' denotes a GPU out-of-memory error.)}}
  \label{tab:impact:batching}
\end{table}

We explore the benefit of the execution phase in \sys{}'s \emph{execute/agree/attest} strategy under different workloads.

\begin{figure}[tb]
  \centering
  \includegraphics[width=1.0\columnwidth]{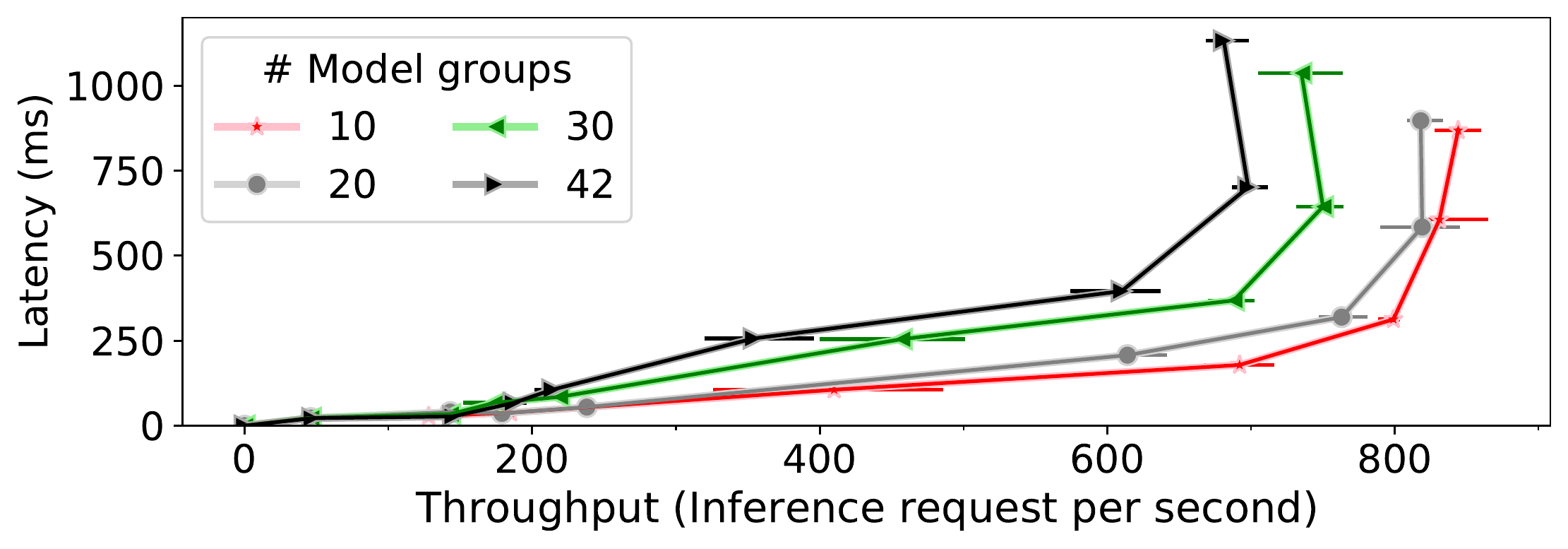}
  \caption{Varying model group count \textnormal{($f$$=$$1$ on \textsf{cloud/single-site}.)}}
  \label{fig:throughput-vs-latency-model-variance}
\end{figure}

First, we measure throughput and latency with different numbers of model groups (10, 20, 30, 42) over which inference requests are served. \Cref{fig:throughput-vs-latency-model-variance} shows that \sys achieves a throughput of 831 and 698\unit{requests/s} with 10~model groups and 42~model groups, respectively, while the latency stays below 700\unit{ms}. The throughput with more model groups decreases because it becomes easier to create large execution batches that refer to the same model.

We confirm this explanation by investigating how the maximum execution batch size and the model group count affects throughput. We consider our two request workloads, \textsf{inference} and \textsf{inference+updates}. \Cref{tab:impact:batching} shows that, with the \textsf{inference+updates} workload, batch sizes of 4 and 8 do not affect throughput, but a batch size of 4 allows \sys to support more loaded model groups due to GPU memory constraints. For that reason, we use a batch size of 4 in our experiments for \textsf{cloud/single-site}.

With the \textsf{inference} workload, throughput improves when increasing the execution batch size. This result is consistent with findings for other inference services~\cite{gujarati2020serving, romero2021infaas, tritonin57:online}. The impact of loading models means that the PCIe bus becomes saturated, delaying the transfer of requests to the GPU, which mitigates the benefit of larger execution batches.

Second, we examine the impact of separating execution and agreement batches. For this, we implement a \sys variant that executes inference requests during the agreement phase of the Byzantine consensus protocol (\textsf{agree/execute}). In this approach, the consensus protocol first orders requests into agreement batches and then executes them.

\begin{figure}[tb]
  \centering
  \includegraphics[width=1.0\columnwidth]{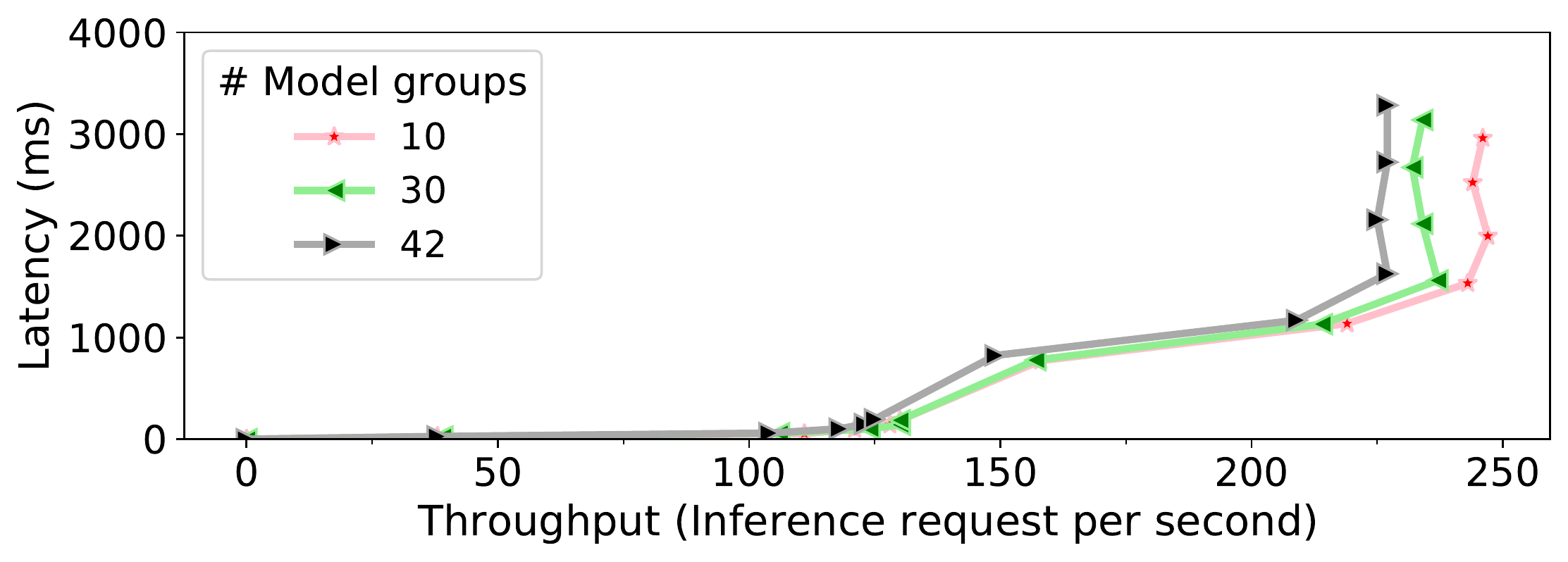}
  \caption{With \textsf{agree/execute} strategy \textnormal{($f$$=$$1$ on \textsf{cloud/single-site}.)}}
  \label{fig:naive-impl}
\end{figure}

\Cref{fig:naive-impl} shows a throughput vs.\ latency plot of \sys with \textsf{agree/execute} for different model group numbers. Here the peak throughput is 246\unit{request/s}, compared to \sys's 854\unit{requests/s}. This 3$\times$ performance reduction stems from two issues: (i) the \textsf{agree/execute} strategy results in lower GPU utilization. While it can exploit larger agreement batches (3 versus 2) to mitigate this, this still fails to remove all GPU idle periods. Even when we turn off execution batches, \sys still achieves 316\unit{requests/s}; and (ii)~the lack of batching during inference execution with \textsf{agree/execute}. When inference requests are executed out-of-order within an agreement batch, there are fewer opportunities to batch inference requests, because agreement batches contain fewer executable requests (25 in \textsf{cloud/single-site}; 50 in \textsf{cloud/multi-site}). In contrast, with the \emph{execute/agree/attest} strategy, the execution batch size is bounded only by the number of concurrent inference requests submitted by clients.

\subsection{Model replication vs.\ agreement}
\label{subsec:agg-vs-rep}

We consider the overhead that \sys{}'s agreement approach using a distance function introduces. We compare against a variant of \sys that only replicates the same sample model on multiple nodes without agreement.

\begin{figure}[tb]
  \centering
  \includegraphics[width=1.0\columnwidth]{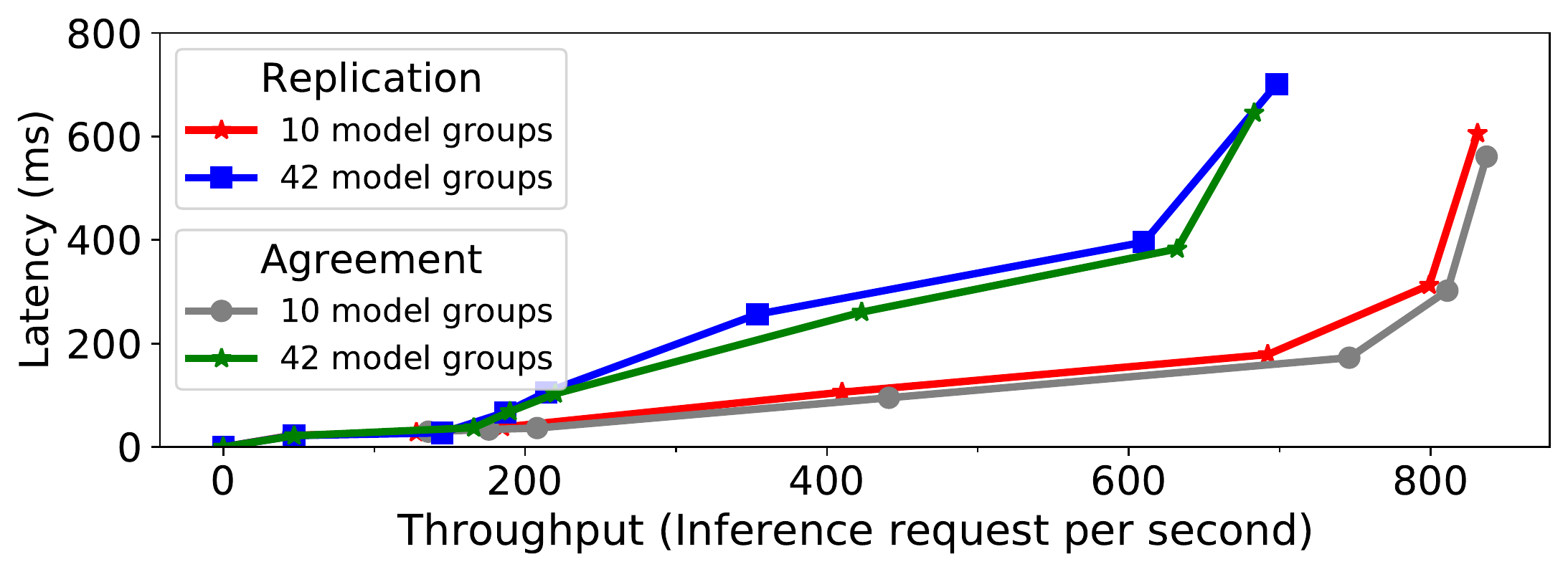}
  \caption{Model replication vs. agreement \textnormal{($f$$=$$1$ on \textsf{cloud/single-site}.)}}
  \label{fig:throughput-vs-latency}
\end{figure}

\Cref{fig:throughput-vs-latency} shows a throughput/latency plot when the models in the model groups are either replicated or use agreement. With replication, we see a maximum throughput of 837\unit{requests/s} with 10~model groups and 683\unit{requests/s} with 42~model groups with a latency of 561\unit{ms} and 646\unit{ms}, respectively; using agreement obtains a peak throughput of 831 and 698\unit{requests/s} with 10~model groups and 42~model groups, with latency of 606\unit{ms} and 700\unit{ms}, respectively. Agreement increases latency by 8\% under heavy client load; under low load, latency increases only by 2\%. This is explained by the computational requirements of the distance function.

\subsection{Updating model groups}
\label{subsec:model-update}

We explore how model updates affect concurrently executing inference requests. In this experiment, we slowly increase the percentage of model groups that are concurrently updated. We consider two configurations with 10 and 30 model groups.

\begin{figure}[tb]
  \centering
    \includegraphics[width=1.0\columnwidth]{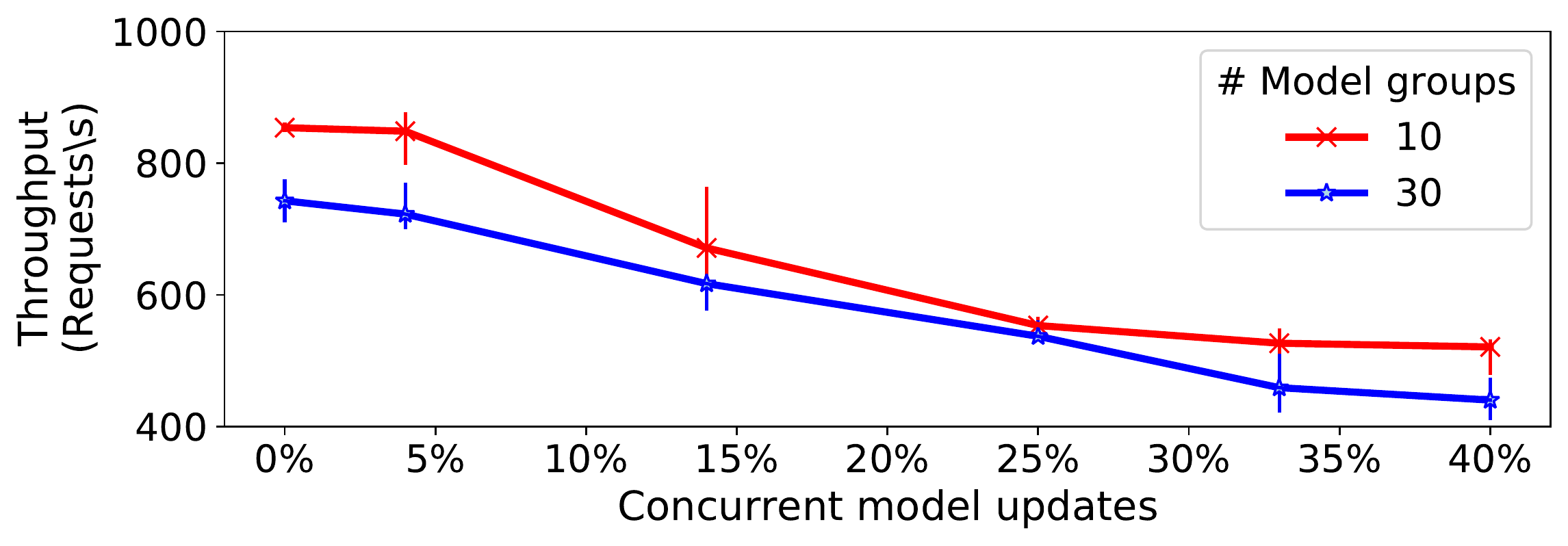}
  \caption{Concurrent model updates \textnormal{($f$$=$$1$ on \textsf{cloud/single-site}.)}}
  \label{fig:model-update-Throughput}
\end{figure}

\Cref{fig:model-update-Throughput} shows that, with 30~model groups, there is a gentle decline in the number of executed inference requests as the percentage of model group updates increases. This is due to three factors, which grow with the number of concurrent updates: (i)~the network bandwidth used to copy models from remote storage; (ii)~the PCIe utilization when models are copied to the GPU; and (iii)~the overhead of executing inference requests against multiple models when a new model has been loaded but not yet activated. These three factors reduce throughput from 743\unit{requests/s}, when no model groups are updated, to 440\unit{requests/s} with 40\% of the model groups concurrently updated. When more than 34\% of the model groups are updated, the updates begin queueing at the GPUs. This reduces the impact on inference requests, as can be seen from the changed slope.

With fewer model groups~(10), there is a faster degradation in the inference request rate until 25\% model groups are concurrently updated. This is due to the greater likelihood of inference requests being executed against an updating model group, thus requiring to execute multiple inference requests against the current and new model. This decline abates when 34\% of model groups are concurrently updated due to queuing, as \sys only allows one concurrent update per group.

\subsection{Scalability}
\label{subsec:scalability}

\begin{figure}[tb]
  \centering
  \begin{subfigure}[t]{.49\linewidth}
    \centering
    \includegraphics[width=1.0\columnwidth]{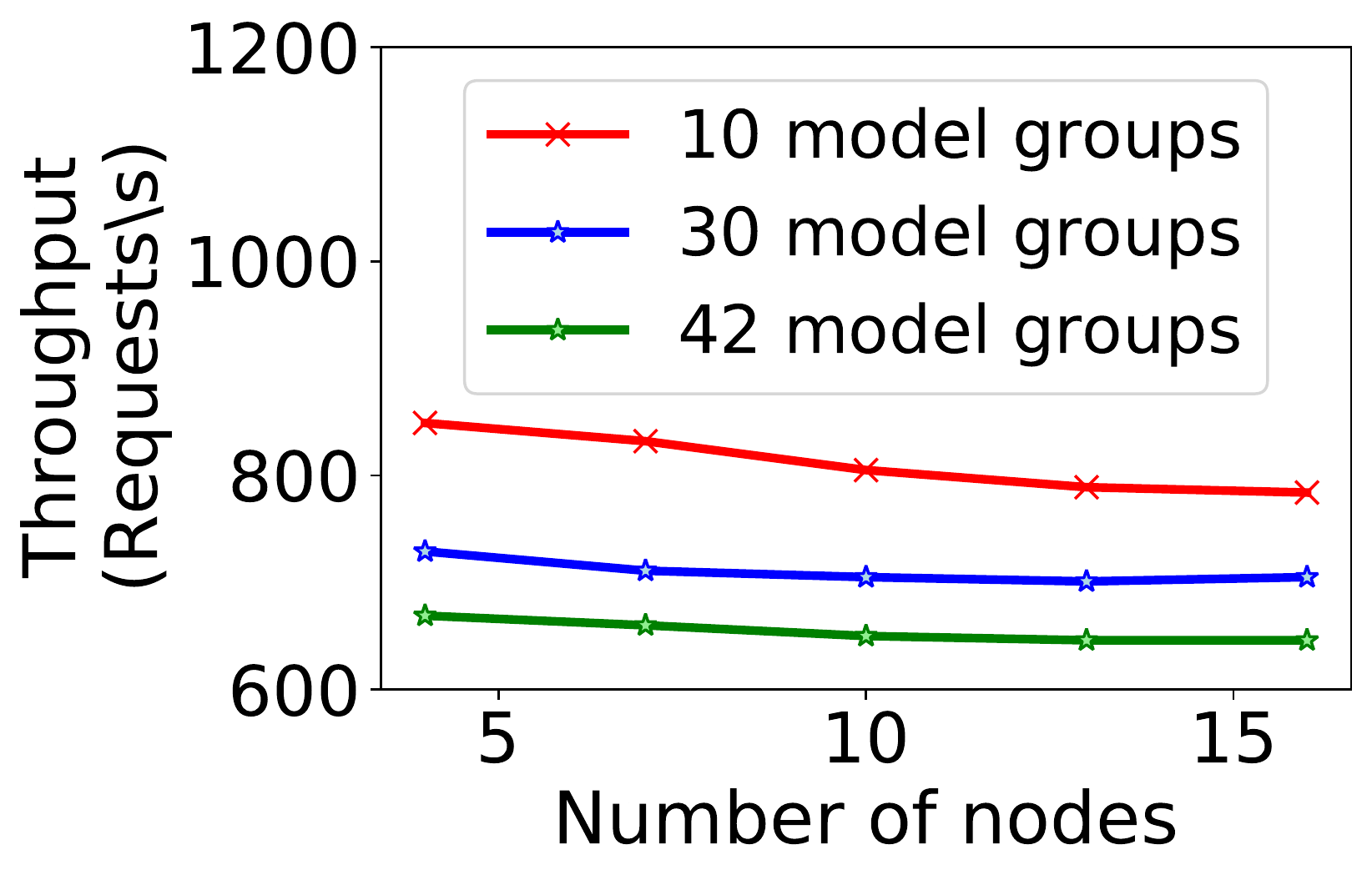}
    \caption{inference}\label{fig:scale-out:ensembles}
  \end{subfigure}
  \begin{subfigure}[t]{.49\linewidth}
    \centering
    \includegraphics[width=1.0\columnwidth]{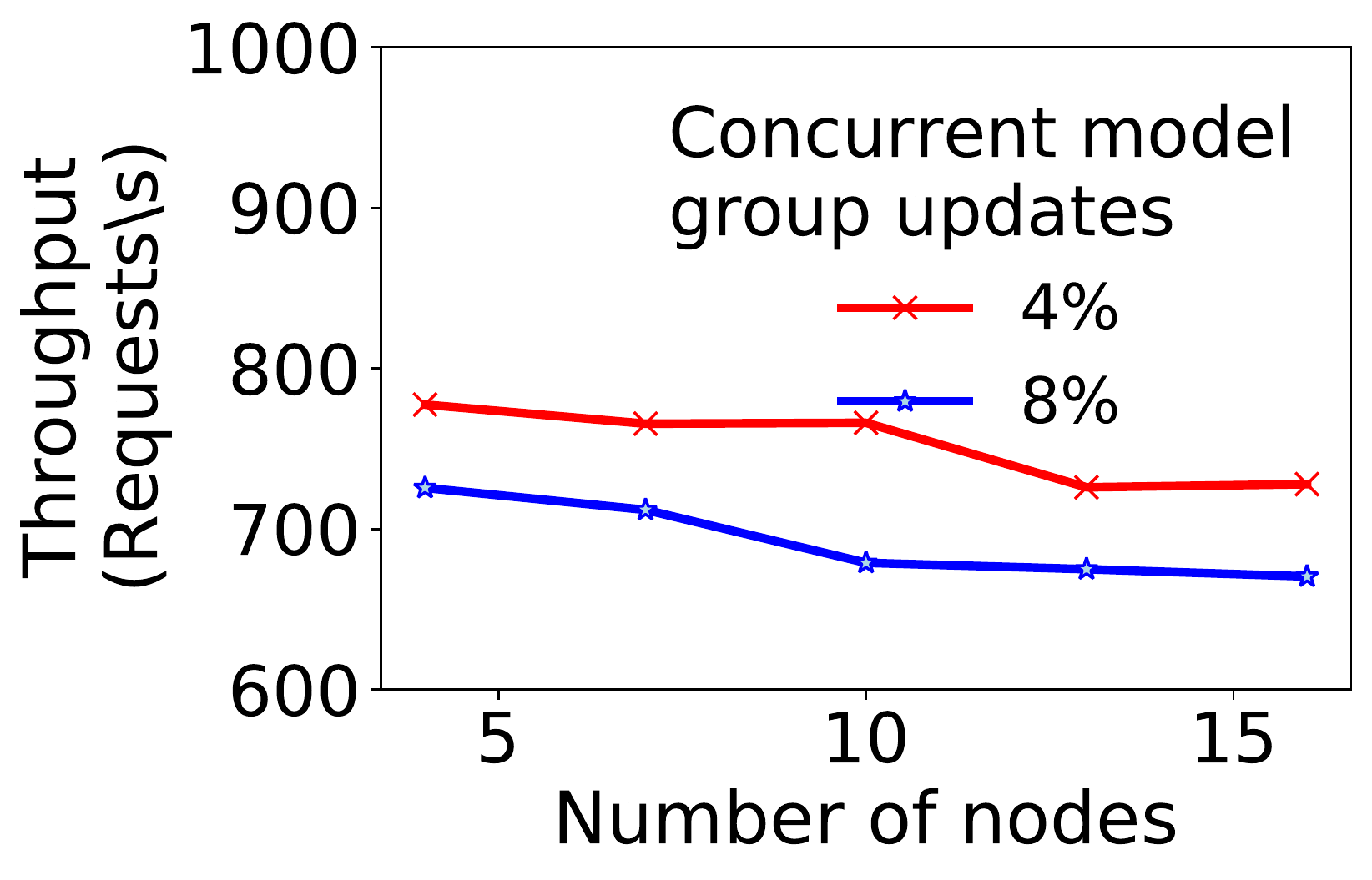}
    \caption{inference+update}  \label{fig:scale-out:updates}
    \end{subfigure}
  \caption{Increasing node count \textnormal{($f$$=$$1$ to $5$ on \textsf{cloud/multi-site}.)}}
  \label{fig:scale-out}
\end{figure}

\sys can scale linearly by load-balancing requests across multiple independent deployments. This has no impact on clients, because inference request execution is stateless on the nodes. Each model group must be assigned to one deployment, otherwise the serializability of model group updates is not guaranteed. A deployment can also be scaled up by using larger VMs sizes, \eg with more GPUs.

In a single \sys deployment, we can mitigate against more untrustworthy nodes~$f$ by adding more nodes. We conduct an experiment that explores how scalable \sys{}'s design is with more nodes. We use our \textsf{cloud/multi-site} configuration across 3~Azure regions and change the node count from 4 to 16, \ie increasing $f$ from $1$ to $5$. We place the primary node in US South Central, forcing worst-case round-trip network latencies of 32\unit{ms}.

First, we explore how the throughput with different model group counts is affected when increasing node count. \Cref{fig:scale-out:ensembles} shows that, with 4~nodes distributed across the 3~Azure regions, the throughput of 849\unit{requests/s} is similar to the previous \textsf{cloud/single-site} experiment~(see~\Cref{fig:throughput-vs-latency-model-variance}). With 16~nodes, throughput degrades to 784\unit{requests/s} when 10~model groups are loaded. Adding more nodes increases the message and cryptographic overheads due to the extra signatures and Merkle tree verification in every agreement batch. The throughput degradation with 30 and 42 loaded model groups is similar. The reduced throughput with more model groups is also caused by the less effective batching---the same effect as in the \textsf{cloud/single-site} experiments (see \S\ref{subsec:impact-batching}).

Next, we investigate how the workload mix of inference and update requests affects throughput with more nodes. \Cref{fig:scale-out:updates} shows \sys throughput in the \textsf{cloud/multi-site} configuration with clients that issue \texttt{define\_model\_group} and \texttt{activate\_model\_group} requests. When 4\% of model groups are updated, inference requests decrease from 777 to 728\unit{requests/s}, as $f$ increases from 1 to 5; the throughput decreases from 726 to 670\unit{requests/s} when 10\% of the model groups are updated. Increasing the number of concurrent model group updates results in higher PCIe utilization, forcing inference requests to queue when transferred to the GPU.

\subsection{Untrustworthy nodes}
\label{subsec:impact-failure}

\begin{table}[tb]
\centering
  \footnotesize\addtolength{\tabcolsep}{-3pt}
  \begin{subtable}[t]{.49\linewidth}
    \centering
    \begin{tabular}{ lc } 
      \toprule  
      \textbf{Dishonest result} & \textbf{Accuracy} \\
      \midrule
      VGG19        & 76.9\% \\
      DenseNet-201 & 75.9\%  \\
      ResNet152    & 75.8\% \\
      ResNeXt101   & 75.3\%  \\
      \bottomrule
    \end{tabular}
    \caption{\textnormal{model group A}}
  \end{subtable}
  \begin{subtable}[t]{.49\linewidth}
    \centering
    \begin{tabular}{ lc } 
      \toprule  
      \textbf{Dishonest result} & \textbf{Accuracy} \\
      \midrule
      VGG11        & 66.3\%  \\
      ResNet18     & 65.4\%  \\
      GoogleNet    & 65.1\% \\
      MnasNet      & 63.0\% \\
      \bottomrule
    \end{tabular}
    \caption{\textnormal{model group B}}
  \end{subtable}
  \caption{Inference accuracy with untrustworthy nodes \normalfont{(Top-1 accuracy when 1 node returns a dishonest result. ($f$$=$$1$)}}%
  \label{tab:ensemble:results}
\end{table}

\begin{figure}[tb]
  \centering
  \includegraphics[width=1.0\columnwidth]{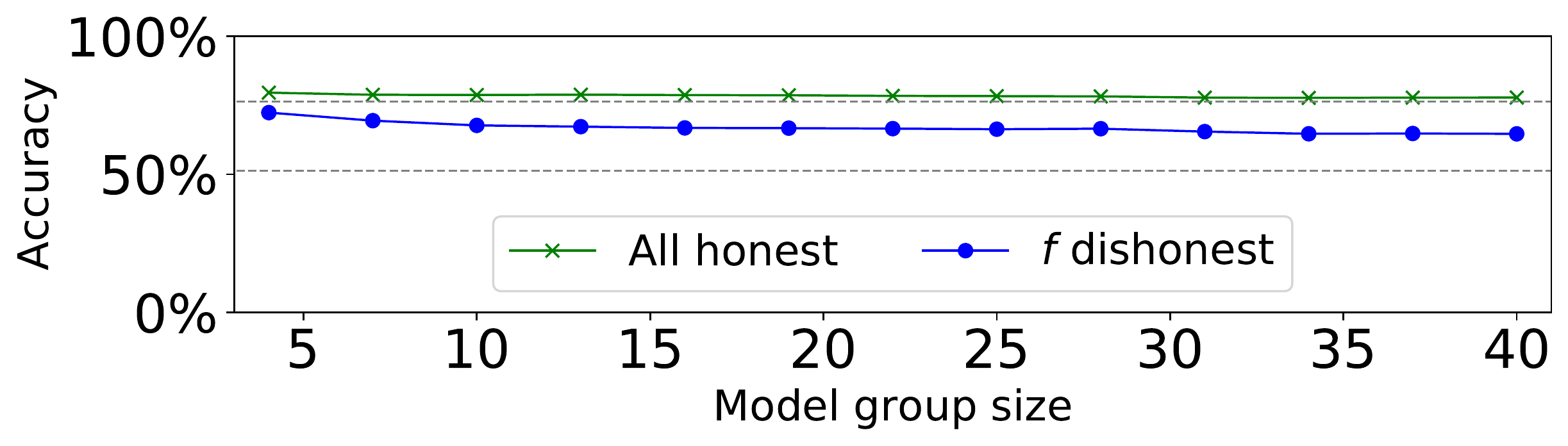}
  \caption{Accuracy versus model group size \textnormal{(Min/max single model accuracy shown as grey dashed lines.)}}
  \label{fig:accuracy-inc-f}
\end{figure}

We examine how untrustworthy nodes affect the accuracy of an inference result in a model group. For this, we use the ImageNet1000 (mini) dataset~\cite{deng2009imagenet} with 2~model groups, each with 4 models: model group~A contains ResNet152, VGG19, DenseNet-201, and ResNeXt101; and model group~B contains GoogleNet, ResNet18, MnasNet0.5, and VGG11.

We combine the results from inference requests: we take the label with the highest confidence and check if more than $f$~models made the same prediction. When all nodes are trustworthy, model group~A has an accuracy of 79.9\% and model group~B has an accuracy 70.0\%. Here the model group improves accuracy up to 7.5pp compared to a single model.

We now consider when an untrustworthy node returns a dishonest result. As shown in \Cref{tab:ensemble:results}, with one untrustworthy node, model group~A achieves a maximum accuracy of 76.9\% and minimum accuracy of 75.3\%---a reduction by up to 4.6pp; model group~B has an accuracy range of 63.0\% to 66.3\% (reduced by up to 7pp). With the evaluated models, an untrustworthy node reduces accuracy by up to 2pp compared to replicating a single model. We observe that, in this scenario, \sys exhibits an accuracy improvement over replication with only honest results and only a small accuracy reduction with a dishonest result.

We also examine the accuracy as the number of models in a model group increases (see \Cref{fig:accuracy-inc-f}). The figure shows two data series: in the first, all inference results are trustworthy. Here, accuracy gently decreases with the number of models and is always higher than the accuracy of the most accurate model. This is caused by the fact that more results must agree on the top inference result; the second line shows the accuracy when $f$ inference results are untrustworthy, and we notice that accuracy decreases at a faster rate. We assume a worst case scenario in which the dishonest inference results are from models that would have returned a correct prediction. Even in this case, the accuracy is still higher than that of the model with the lowest accuracy.

\subsection{Size of inference requests and results}
\label{subsec:impact-req-res-size}

\begin{figure}[tb]
  \centering
  \begin{subfigure}[t]{.49\linewidth}
    \centering
    \includegraphics[width=1.0\columnwidth]{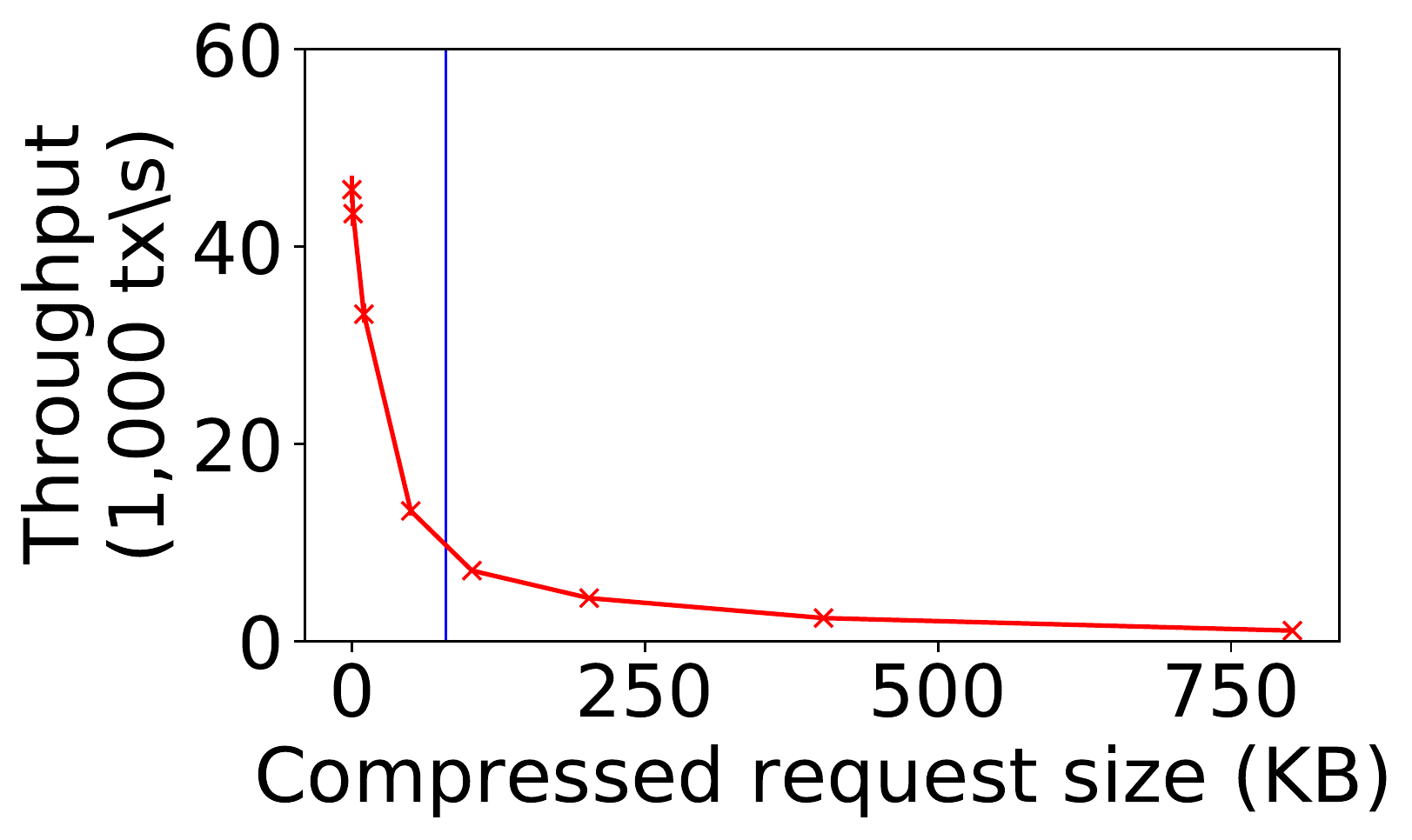}
    \caption{Increasing request size}\label{fig:scale-out-response-size:request}
  \end{subfigure}
  \begin{subfigure}[t]{.49\linewidth}
    \centering
    \includegraphics[width=1.0\columnwidth]{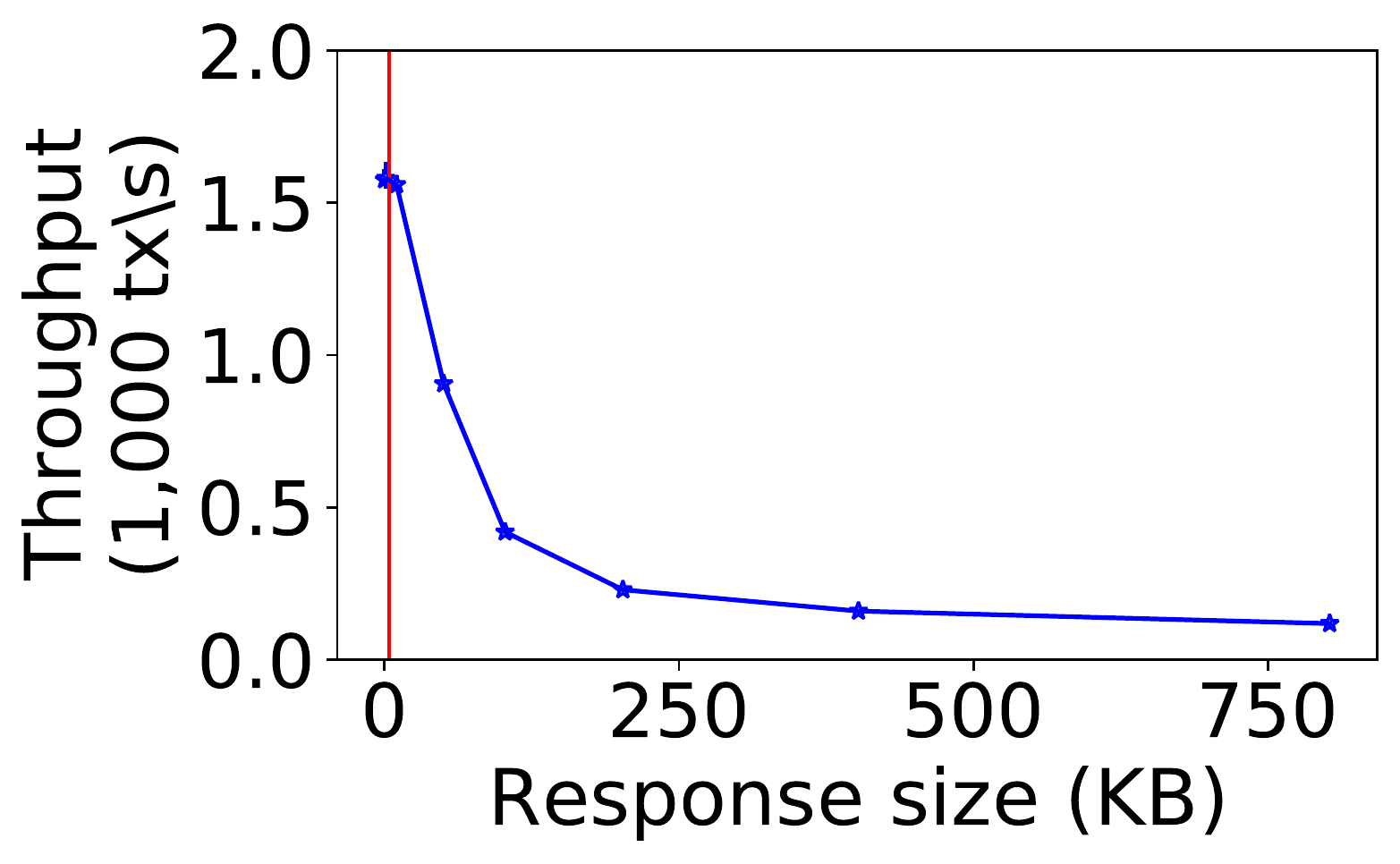}
    \caption{Increasing response size}\label{fig:scale-out-response-size:result}
    \end{subfigure}
    \caption{Inference request/result size \normalfont{($f$$=$$1$ on \textsf{cluster}.)}}
  \label{fig:scale-out-response-size}
\end{figure}

To understand how \sys is impacted by the size of requests and responses, we measure throughput after replacing GPU inference execution with a randomly sampled response. Generally, models that operate on larger inputs/outputs are more complex and thus would be GPU bottlenecked~(see \S\ref{subsec:comparison}).

First, we vary the compressed request size from 16\unit{bytes} to 800\unit{KB}, with a fixed response size of 4000\unit{bytes}. Note that \sys clients compress images using the lowest level of zlib~\cite{deutsch1996zlib} compression, and we find that the average compressed image size from ImageNet1000 (mini)~\cite{deng2009imagenet} is 80\unit{KB}.

\Cref{fig:scale-out-response-size:request} shows that a peak throughput of 45,765\unit{requests/s} is achieved with 16-byte requests. With 1\unit{KB} and 10\unit{KB} requests, throughput reduces to 43,335\unit{requests/s} and 33,149\unit{requests/s}, respectively, with larger requests halving throughput as the request size doubles. The typical ImageNet size of 80\unit{KB} (blue line) achieves 1581\unit{requests/s}.

On small inputs, the overhead of verifying inference request signatures becomes the bottleneck. When compressed requests are larger than 100\unit{KB}, the bottleneck shifts to creating and verifying the Merkle trees over the requests/responses~(see \S\ref{sec:design:agree}). This cost could be reduced by splitting the inputs into smaller chunks in the Merkle tree and computing their hashes in parallel across a larger number of threads.

In \Cref{fig:scale-out-response-size:result}, we vary the inference result from 16\unit{bytes} to 800\unit{KB}, with the representative compressed request size of 80~\unit{KB}. We observe that peak throughput is obtained when results are between 16\unit{bytes} and 10\unit{KB}, with typical ResNet output of 4\unit{KB} marked as a red line. For results over 50\unit{KB}, throughput halves as the size doubles. Similar to the previous experiment, \sys remains CPU bound on signature verification when the results are small, and the bottleneck shifts to Merkle tree computation for result sizes larger than 10\unit{KB}.


%% file: 7-related-work.tex
\section{Related Work}
\label{sec:rel_work}

\mypar{Distributed ML inference} INFaaS~\cite{romero2021infaas} is an ML inference system that selects an appropriate model based on a confidence level and time budget. Clockwork~\cite{gujarati2020serving} makes the latency of inference execution predictable. Clipper~\cite{crankshaw2017clipper} provides an abstraction layer on top of multiple inference frameworks. None of these systems assume untrustworthy nodes, which may tamper with the returned results. Model-Switch~\cite{zhang2020model} selects models that reduce inference confidence to maintain a request latency SLO. Tolerance Tiers~\cite{halpern2019one} provides an API that allows users to decide if a simpler model should be used when an SLO is violated. \sys offers trustworthy inference, but it could use SLO-aware policies for model selection from an ensemble.

\mypar{Secure ML inference} Securely obtaining inference results has been studied before. Bost \etal\cite{bost2014machine} and SecureML~\cite{mohassel2017secureml} use privacy-preserving techniques to execute particular ML inference workloads, but they cannot provide a general trustworthy ML inference service in the cloud. CrypTFlow~\cite{kumar2020cryptflow} and CrypTFlow2~\cite{crankshaw2017clipper} obtain cryptographically secure inference results through multi-party computation. While they protect larger models, such as ResNet, their performance precludes practical deployments: they require over a minute to compute a single ResNet50 inference result.

Slalom~\cite{tramer2018slalom} and Privado~\cite{grover2018privado} use trusted execution environment~(TEEs)~\cite{costan2016intel} to produce trustworthy inference. They rely on the security of TEEs, which have suffered from successful attacks~\cite{buhren2021one, murdock2020plundervolt, van2020lvi, nilsson2020survey}. These solutions are restricted to CPU-based inference, which increases latencies by several orders of magnitude compared to GPUs. Graviton~\cite{volos2018graviton} proposes a novel TEE abstraction that encompasses a GPU, but it requires specific GPU hardware support.

\mypar{Byzantine ML} To our knowledge, \sys is the first ML inference system that assumes Byzantine node behaviour. Byzantine failures have been shown to be catastrophic for ML training~\cite{guerraoui2018hidden}, and past work has investigated the impact of Byzantine adversaries on distributed optimization~\cite{data2020data}. Krum~\cite{blanchard2017machine} and trimmed mean~\cite{yin2018byzantine} create Byzantine-resilient algorithms for distributed stochastic gradient descent~(SGD), providing mechanisms to protect against some Byzantine failures. Such approaches can be combined with \sys to mitigate malicious nodes during training.


%% file: 8-conclusion.tex
\section{Conclusion}
\label{sec:concl}

Trustworthy inference results allow clients to rely on cloud-based ML inference services. \sys is the first design for a trustworthy cloud-based inference service, which provides clients with an inference certificate to prove that results are trustworthy. \sys uses an execute/agree/attest strategy, which allows trustworthy inference results to be produced with a negligible performance overhead---it separates the execution of requests on GPUs from their ordering in agreement batches using a Byzantine fault-tolerant consensus protocol.


%% file: paper.bbl
\begin{thebibliography}{100}

\bibitem{aggour2006automating}
{\sc Aggour, K.~S., Bonissone, P.~P., Cheetham, W.~E., and Messmer, R.~P.}
\newblock Automating the underwriting of insurance applications.
\newblock {\em AI magazine 27}, 3 (2006), 36--36.

\bibitem{AmazonSa44:online}
Amazon sagemaker – machine learning – amazon web services.
\newblock \url{https://aws.amazon.com/sagemaker/}.
\newblock (Accessed on 07/28/2021).

\bibitem{OutOfThe46:online}
Out of the box ai solutions | antal.ai - zsiros antal computer vision
  specialist.
\newblock \url{https://www.antal.ai/}.
\newblock (Accessed on 04/16/2022).

\bibitem{onnx}
{\sc Bai, J., Lu, F., Zhang, K., et~al.}
\newblock Onnx: Open neural network exchange.
\newblock \url{https://github.com/onnx/onnx}.

\bibitem{bessani2014state}
{\sc Bessani, A., Sousa, J., and Alchieri, E.~E.}
\newblock State machine replication for the masses with bft-smart.
\newblock In {\em 2014 44th Annual IEEE/IFIP International Conference on
  Dependable Systems and Networks\/} (2014), IEEE, pp.~355--362.

\bibitem{BigVisio20:online}
Big vision | consulting services in ai, computer vision, and deep learning.
\newblock \url{https://bigvision.ai/}.
\newblock (Accessed on 04/16/2022).

\bibitem{biggio2011bagging}
{\sc Biggio, B., Corona, I., Fumera, G., Giacinto, G., and Roli, F.}
\newblock Bagging classifiers for fighting poisoning attacks in adversarial
  classification tasks.
\newblock In {\em International workshop on multiple classifier systems\/}
  (2011), Springer, pp.~350--359.

\bibitem{biggio2012poisoning}
{\sc Biggio, B., Nelson, B., and Laskov, P.}
\newblock Poisoning attacks against support vector machines.
\newblock {\em arXiv preprint arXiv:1206.6389\/} (2012).

\bibitem{blanchard2017machine}
{\sc Blanchard, P., El~Mhamdi, E.~M., Guerraoui, R., and Stainer, J.}
\newblock Machine learning with adversaries: Byzantine tolerant gradient
  descent.
\newblock In {\em Proceedings of the 31st International Conference on Neural
  Information Processing Systems\/} (2017), pp.~118--128.

\bibitem{bost2014machine}
{\sc Bost, R., Popa, R.~A., Tu, S., and Goldwasser, S.}
\newblock Machine learning classification over encrypted data.
\newblock {\em Cryptology ePrint Archive\/} (2014).

\bibitem{breiman1996bagging}
{\sc Breiman, L.}
\newblock Bagging predictors.
\newblock {\em Machine learning 24}, 2 (1996), 123--140.

\bibitem{buhren2021one}
{\sc Buhren, R., Jacob, H.-N., Krachenfels, T., and Seifert, J.-P.}
\newblock One glitch to rule them all: Fault injection attacks against amd's
  secure encrypted virtualization.
\newblock {\em arXiv preprint arXiv:2108.04575\/} (2021).

\bibitem{carbonneau2008application}
{\sc Carbonneau, R., Laframboise, K., and Vahidov, R.}
\newblock Application of machine learning techniques for supply chain demand
  forecasting.
\newblock {\em European Journal of Operational Research 184}, 3 (2008),
  1140--1154.

\bibitem{castro2001practical}
{\sc Castro, M.}
\newblock Practical byzantine fault tolerance, 2001.

\bibitem{pbft}
{\sc Castro, M., and Liskov, B.}
\newblock {Practical Byzantine Fault Tolerance}.
\newblock In {\em OSDI\/} (1999), vol.~99, pp.~173--186.

\bibitem{catania_keefer_1987}
{\sc Catania, P.~J., and Keefer, N.}
\newblock The marketplace, 1987.

\bibitem{chen2018tvm}
{\sc Chen, T., Moreau, T., Jiang, Z., Shen, H., Yan, E.~Q., Wang, L., Hu, Y.,
  Ceze, L., Guestrin, C., and Krishnamurthy, A.}
\newblock Tvm: end-to-end optimization stack for deep learning.
\newblock {\em arXiv preprint arXiv:1802.04799 11\/} (2018), 20.

\bibitem{choi2018generative}
{\sc Choi, H., and Jang, E.}
\newblock Generative ensembles for robust anomaly detection, 2018.

\bibitem{clement2009making}
{\sc Clement, A., Wong, E.~L., Alvisi, L., Dahlin, M., and Marchetti, M.}
\newblock Making byzantine fault tolerant systems tolerate byzantine faults.
\newblock In {\em NSDI\/} (2009), vol.~9, pp.~153--168.

\bibitem{coglianese2020ai}
{\sc Coglianese, C., and Ben~Dor, L.}
\newblock Ai in adjudication and administration.
\newblock {\em Brooklyn Law Review, Forthcoming, U of Penn Law School, Public
  Law Research Paper}, 19-41 (2020).

\bibitem{TheHidde96:online}
| the hidden consequences of moderating social media’s dark side.
\newblock
  \url{https://contentmarketinginstitute.com/cco-digital/july-2019/social-media-moderators-stress/}.
\newblock (Accessed on 04/17/2022).

\bibitem{costan2016intel}
{\sc Costan, V., and Devadas, S.}
\newblock Intel sgx explained.
\newblock {\em IACR Cryptol. ePrint Arch. 2016}, 86 (2016), 1--118.

\bibitem{crankshaw2017clipper}
{\sc Crankshaw, D., Wang, X., Zhou, G., Franklin, M.~J., Gonzalez, J.~E., and
  Stoica, I.}
\newblock Clipper: A low-latency online prediction serving system.
\newblock In {\em 14th $\{$USENIX$\}$ Symposium on Networked Systems Design and
  Implementation ($\{$NSDI$\}$ 17)\/} (2017), pp.~613--627.

\bibitem{cyphers2018intel}
{\sc Cyphers, S., Bansal, A.~K., Bhiwandiwalla, A., Bobba, J., Brookhart, M.,
  Chakraborty, A., Constable, W., Convey, C., Cook, L., Kanawi, O., et~al.}
\newblock Intel ngraph: An intermediate representation, compiler, and executor
  for deep learning.
\newblock {\em arXiv preprint arXiv:1801.08058\/} (2018).

\bibitem{dalvi2004adversarial}
{\sc Dalvi, N., Domingos, P., Sanghai, S., and Verma, D.}
\newblock Adversarial classification.
\newblock In {\em Proceedings of the tenth ACM SIGKDD international conference
  on Knowledge discovery and data mining\/} (2004), pp.~99--108.

\bibitem{data2020data}
{\sc Data, D., Song, L., and Diggavi, S.~N.}
\newblock Data encoding for byzantine-resilient distributed optimization.
\newblock {\em IEEE Transactions on Information Theory 67}, 2 (2020),
  1117--1140.

\bibitem{debreczeni2019neural}
{\sc Debreczeni, G.}
\newblock Neural network exchange format, 2019.

\bibitem{deng2009imagenet}
{\sc Deng, J., Dong, W., Socher, R., Li, L.-J., Li, K., and Fei-Fei, L.}
\newblock Imagenet: A large-scale hierarchical image database.
\newblock In {\em 2009 IEEE conference on computer vision and pattern
  recognition\/} (2009), Ieee, pp.~248--255.

\bibitem{deutsch1996zlib}
{\sc Deutsch, P., and Gailly, J.-L.}
\newblock Zlib compressed data format specification version 3.3.
\newblock Tech. rep., RFC 1950, May, 1996.

\bibitem{onnx_runtime}
{\sc developers, O.~R.}
\newblock Onnx runtime.
\newblock \url{https://www.onnxruntime.ai}, 2021.

\bibitem{ContentM17:online}
Content moderation is broken. let us count the ways. | electronic frontier
  foundation.
\newblock
  \url{https://www.eff.org/deeplinks/2019/04/content-moderation-broken-let-us-count-ways}.
\newblock (Accessed on 04/17/2022).

\bibitem{elmrabit2020evaluation}
{\sc Elmrabit, N., Zhou, F., Li, F., and Zhou, H.}
\newblock Evaluation of machine learning algorithms for anomaly detection.
\newblock In {\em 2020 International Conference on Cyber Security and
  Protection of Digital Services (Cyber Security)\/} (2020), IEEE, pp.~1--8.

\bibitem{eykholt2018robust}
{\sc Eykholt, K., Evtimov, I., Fernandes, E., Li, B., Rahmati, A., Xiao, C.,
  Prakash, A., Kohno, T., and Song, D.}
\newblock Robust physical-world attacks on deep learning visual classification.
\newblock In {\em Proceedings of the IEEE conference on computer vision and
  pattern recognition\/} (2018), pp.~1625--1634.

\bibitem{gao2019ai}
{\sc Gao, J., Tao, C., Jie, D., and Lu, S.}
\newblock What is ai software testing? and why.
\newblock In {\em 2019 IEEE International Conference on Service-Oriented System
  Engineering (SOSE)\/} (2019), IEEE, pp.~27--2709.

\bibitem{GartnerI61:online}
Gartner identifies four trends driving near-term artificial intelligence
  innovation.
\newblock
  \url{https://www.gartner.com/en/newsroom/press-releases/2021-09-07-gartner-identifies-four-trends-driving-near-term-artificial-intelligence-innovation}.
\newblock (Accessed on 11/04/2021).

\bibitem{gillespie2020content}
{\sc Gillespie, T.}
\newblock Content moderation, ai, and the question of scale.
\newblock {\em Big Data \& Society 7}, 2 (2020), 2053951720943234.

\bibitem{Whatisth51:online}
{\sc Gold, R.~N.}
\newblock What is the impact of machine learning on mortgage lending?
\newblock
  \url{https://wrds-www.wharton.upenn.edu/pages/news/what-impact-machine-learning-mortgage-lending/}.
\newblock (Accessed on 04/19/2022).

\bibitem{Goodfellow-et-al-2016}
{\sc Goodfellow, I., Bengio, Y., and Courville, A.}
\newblock {\em Deep Learning}.
\newblock MIT Press, 2016.
\newblock \url{http://www.deeplearningbook.org}.

\bibitem{CloudInf6:online}
Cloud inference api | cloud inference api | google cloud.
\newblock \url{https://cloud.google.com/inference}.
\newblock (Accessed on 07/28/2021).

\bibitem{google_ai_ml}
Ai \& machine learning products | google cloud.

\bibitem{gorwa2020algorithmic}
{\sc Gorwa, R., Binns, R., and Katzenbach, C.}
\newblock Algorithmic content moderation: Technical and political challenges in
  the automation of platform governance.
\newblock {\em Big Data \& Society 7}, 1 (2020), 2053951719897945.

\bibitem{grover2018privado}
{\sc Grover, K., Tople, S., Shinde, S., Bhagwan, R., and Ramjee, R.}
\newblock Privado: Practical and secure dnn inference with enclaves.
\newblock {\em arXiv preprint arXiv:1810.00602\/} (2018).

\bibitem{guerraoui2018hidden}
{\sc Guerraoui, R., Rouault, S., et~al.}
\newblock The hidden vulnerability of distributed learning in byzantium.
\newblock In {\em International Conference on Machine Learning\/} (2018), PMLR,
  pp.~3521--3530.

\bibitem{gujarati2020serving}
{\sc Gujarati, A., Karimi, R., Alzayat, S., Hao, W., Kaufmann, A., Vigfusson,
  Y., and Mace, J.}
\newblock Serving dnns like clockwork: Performance predictability from the
  bottom up.
\newblock In {\em 14th $\{$USENIX$\}$ Symposium on Operating Systems Design and
  Implementation ($\{$OSDI$\}$ 20)\/} (2020), pp.~443--462.

\bibitem{Sommelier}
{\sc Guo, P., Hu, B., , and Hu, W.}
\newblock Sommelier: Curating dnn models for the masses.
\newblock In {\em SIGMOD\/} (2022).

\bibitem{hall2017artificial}
{\sc Hall, S.}
\newblock How artificial intelligence is changing the insurance industry.
\newblock {\em The Center for Insurance Policy \& Research 22\/} (2017), 1--8.

\bibitem{halpern2019one}
{\sc Halpern, M., Boroujerdian, B., Mummert, T., Duesterwald, E., and Reddi,
  V.~J.}
\newblock One size does not fit all: Quantifying and exposing the
  accuracy-latency trade-off in machine learning cloud service apis via
  tolerance tiers.
\newblock {\em arXiv preprint arXiv:1906.11307\/} (2019).

\bibitem{he2016deep}
{\sc He, K., Zhang, X., Ren, S., and Sun, J.}
\newblock Deep residual learning for image recognition.
\newblock In {\em Proceedings of the IEEE conference on computer vision and
  pattern recognition\/} (2016), pp.~770--778.

\bibitem{he2016identity}
{\sc He, K., Zhang, X., Ren, S., and Sun, J.}
\newblock Identity mappings in deep residual networks.
\newblock In {\em European conference on computer vision\/} (2016), Springer,
  pp.~630--645.

\bibitem{hillman2019use}
{\sc Hillman, N.~L.}
\newblock The use of artificial intelligence in gauging the risk of recidivism.
\newblock {\em The Judges' Journal 58}, 1 (2019), 36--39.

\bibitem{huang2017densely}
{\sc Huang, G., Liu, Z., Van Der~Maaten, L., and Weinberger, K.~Q.}
\newblock Densely connected convolutional networks.
\newblock In {\em Proceedings of the IEEE conference on computer vision and
  pattern recognition\/} (2017), pp.~4700--4708.

\bibitem{huttenlocher1993comparing}
{\sc Huttenlocher, D.~P., Klanderman, G.~A., and Rucklidge, W.~J.}
\newblock Comparing images using the hausdorff distance.
\newblock {\em IEEE Transactions on pattern analysis and machine intelligence
  15}, 9 (1993), 850--863.

\bibitem{iandola2016squeezenet}
{\sc Iandola, F.~N., Han, S., Moskewicz, M.~W., Ashraf, K., Dally, W.~J., and
  Keutzer, K.}
\newblock Squeezenet: Alexnet-level accuracy with 50x fewer parameters and< 0.5
  mb model size.
\newblock {\em arXiv preprint arXiv:1602.07360\/} (2016).

\bibitem{jameel2015automatic}
{\sc Jameel, T., Mengxiang, L., and Chao, L.}
\newblock Automatic test oracle for image processing applications using support
  vector machines.
\newblock In {\em 2015 6th IEEE International Conference on Software
  Engineering and Service Science (ICSESS)\/} (2015), IEEE, pp.~1110--1113.

\bibitem{jia2020intrinsic}
{\sc Jia, J., Cao, X., and Gong, N.~Z.}
\newblock Intrinsic certified robustness of bagging against data poisoning
  attacks.
\newblock {\em arXiv preprint arXiv:2008.04495\/} (2020).

\bibitem{jouppi2017datacenter}
{\sc Jouppi, N.~P., Young, C., Patil, N., Patterson, D., Agrawal, G., Bajwa,
  R., Bates, S., Bhatia, S., Boden, N., Borchers, A., et~al.}
\newblock In-datacenter performance analysis of a tensor processing unit.
\newblock In {\em Proceedings of the 44th annual international symposium on
  computer architecture\/} (2017), pp.~1--12.

\bibitem{khakurel2018rise}
{\sc Khakurel, J., Penzenstadler, B., Porras, J., Knutas, A., and Zhang, W.}
\newblock The rise of artificial intelligence under the lens of sustainability.
\newblock {\em Technologies 6}, 4 (2018), 100.

\bibitem{kim2014inference}
{\sc Kim, M.~K., and Kim, S.~D.}
\newblock Inference-as-a-service: A situation inference service for
  context-aware computing.
\newblock In {\em 2014 International Conference on Smart Computing\/} (2014),
  IEEE, pp.~317--324.

\bibitem{koh2022stronger}
{\sc Koh, P.~W., Steinhardt, J., and Liang, P.}
\newblock Stronger data poisoning attacks break data sanitization defenses.
\newblock {\em Machine Learning 111}, 1 (2022), 1--47.

\bibitem{krittanawong2018rise}
{\sc Krittanawong, C.}
\newblock The rise of artificial intelligence and the uncertain future for
  physicians.
\newblock {\em European journal of internal medicine 48\/} (2018), e13--e14.

\bibitem{krizhevsky2012imagenet}
{\sc Krizhevsky, A., Sutskever, I., and Hinton, G.~E.}
\newblock Imagenet classification with deep convolutional neural networks.
\newblock {\em Advances in neural information processing systems 25\/} (2012),
  1097--1105.

\bibitem{kugler2018ai}
{\sc Kugler, L.}
\newblock Ai judges and juries.
\newblock {\em Communications of the ACM 61}, 12 (2018), 19--21.

\bibitem{kumar2020marketplace}
{\sc Kumar, A., Finley, B., Braud, T., Tarkoma, S., and Hui, P.}
\newblock Marketplace for ai models.
\newblock {\em arXiv preprint arXiv:2003.01593\/} (2020).

\bibitem{kumar2020cryptflow}
{\sc Kumar, N., Rathee, M., Chandran, N., Gupta, D., Rastogi, A., and Sharma,
  R.}
\newblock Cryptflow: Secure tensorflow inference.
\newblock In {\em 2020 IEEE Symposium on Security and Privacy (SP)\/} (2020),
  IEEE, pp.~336--353.

\bibitem{lee2019occlumency}
{\sc Lee, T., Lin, Z., Pushp, S., Li, C., Liu, Y., Lee, Y., Xu, F., Xu, C.,
  Zhang, L., and Song, J.}
\newblock Occlumency: Privacy-preserving remote deep-learning inference using
  sgx.
\newblock In {\em The 25th Annual International Conference on Mobile Computing
  and Networking\/} (2019), pp.~1--17.

\bibitem{liu2017trojaning}
{\sc Liu, Y., Ma, S., Aafer, Y., Lee, W.-C., Zhai, J., Wang, W., and Zhang, X.}
\newblock Trojaning attack on neural networks, 2017.

\bibitem{maier2019transforming}
{\sc Maier, M., Carlotto, H., Sanchez, F., Balogun, S., and Merritt, S.}
\newblock Transforming underwriting in the life insurance industry.
\newblock In {\em Proceedings of the AAAI Conference on Artificial
  Intelligence\/} (2019), vol.~33, pp.~9373--9380.

\bibitem{marchal2019detecting}
{\sc Marchal, S., and Szyller, S.}
\newblock Detecting organized ecommerce fraud using scalable categorical
  clustering.
\newblock In {\em Proceedings of the 35th Annual Computer Security Applications
  Conference\/} (2019), pp.~215--228.

\bibitem{embedtls}
{SSL Library mbed TLS / PolarSSL}.
\newblock \url{https://tls.mbed.org/}.
\newblock (Accessed on 12/09/2020).

\bibitem{mei2015using}
{\sc Mei, S., and Zhu, X.}
\newblock Using machine teaching to identify optimal training-set attacks on
  machine learners.
\newblock In {\em Twenty-Ninth AAAI Conference on Artificial Intelligence\/}
  (2015).

\bibitem{microsof86:online}
microsoft/merklecpp at 9c92bdb396abe89b9d2796ae0f2093b5f75e17c5.
\newblock
  \url{https://github.com/microsoft/merklecpp/tree/9c92bdb396abe89b9d2796ae0f2093b5f75e17c5}.
\newblock (Accessed on 11/13/2021).

\bibitem{AzureMac41:online}
Azure machine learning - ml as a service | microsoft azure.
\newblock
  \url{https://azure.microsoft.com/en-us/services/machine-learning/#features}.
\newblock (Accessed on 07/28/2021).

\bibitem{cognitive_services}
Cognitive services – apis for ai solutions: Microsoft azure.

\bibitem{mivsura2016data}
{\sc Mi{\v{s}}ura, K., and {\v{Z}}agar, M.}
\newblock Data marketplace for internet of things.
\newblock In {\em 2016 International Conference on Smart Systems and
  Technologies (SST)\/} (2016), IEEE, pp.~255--260.

\bibitem{Modelpla60:online}
Modelplace.ai - modelplace.ai.
\newblock \url{https://modelplace.ai/}.
\newblock (Accessed on 04/16/2022).

\bibitem{AIModelM99:online}
Ai model marketplace - modzy.
\newblock \url{https://www.modzy.com/marketplace/}.
\newblock (Accessed on 04/16/2022).

\bibitem{mohassel2017secureml}
{\sc Mohassel, P., and Zhang, Y.}
\newblock Secureml: A system for scalable privacy-preserving machine learning.
\newblock In {\em 2017 IEEE symposium on security and privacy (SP)\/} (2017),
  IEEE, pp.~19--38.

\bibitem{murdock2020plundervolt}
{\sc Murdock, K., Oswald, D., Garcia, F.~D., Van~Bulck, J., Gruss, D., and
  Piessens, F.}
\newblock Plundervolt: Software-based fault injection attacks against intel
  sgx.
\newblock In {\em 2020 IEEE Symposium on Security and Privacy (SP)\/} (2020),
  IEEE, pp.~1466--1482.

\bibitem{narra2019privacy}
{\sc Narra, K.~G., Lin, Z., Wang, Y., Balasubramaniam, K., and Annavaram, M.}
\newblock Privacy-preserving inference in machine learning services using
  trusted execution environments.
\newblock {\em arXiv preprint arXiv:1912.03485\/} (2019).

\bibitem{narra2021origami}
{\sc Narra, K.~G., Lin, Z., Wang, Y., Balasubramanian, K., and Annavaram, M.}
\newblock Origami inference: Private inference using hardware enclaves.
\newblock In {\em 2021 IEEE 14th International Conference on Cloud Computing
  (CLOUD)\/} (2021), IEEE, pp.~78--84.

\bibitem{nassif2021machine}
{\sc Nassif, A.~B., Talib, M.~A., Nasir, Q., and Dakalbab, F.~M.}
\newblock Machine learning for anomaly detection: a systematic review.
\newblock {\em IEEE Access\/} (2021).

\bibitem{natale2020imagining}
{\sc Natale, S., and Ballatore, A.}
\newblock Imagining the thinking machine: Technological myths and the rise of
  artificial intelligence.
\newblock {\em Convergence 26}, 1 (2020), 3--18.

\bibitem{ni2020systematic}
{\sc Ni, D., Xiao, Z., and Lim, M.~K.}
\newblock A systematic review of the research trends of machine learning in
  supply chain management.
\newblock {\em International Journal of Machine Learning and Cybernetics 11}, 7
  (2020), 1463--1482.

\bibitem{nilsson2020survey}
{\sc Nilsson, A., Bideh, P.~N., and Brorsson, J.}
\newblock A survey of published attacks on intel sgx.
\newblock {\em arXiv preprint arXiv:2006.13598\/} (2020).

\bibitem{GitHubon40:online}
Github - onnx/models: A collection of pre-trained, state-of-the-art models in
  the onnx format.
\newblock \url{https://github.com/onnx/models}.
\newblock (Accessed on 10/24/2021).

\bibitem{httpsope2:online}
https://openauthentication.org.
\newblock \url{https://openauthentication.org/}.
\newblock (Accessed on 08/17/2021).

\bibitem{perlman1999overview}
{\sc Perlman, R.}
\newblock An overview of pki trust models.
\newblock {\em IEEE network 13}, 6 (1999), 38--43.

\bibitem{protzenko2020evercrypt}
{\sc Protzenko, J., Parno, B., Fromherz, A., Hawblitzel, C., Polubelova, M.,
  Bhargavan, K., Beurdouche, B., Choi, J., Delignat-Lavaud, A., Fournet, C.,
  et~al.}
\newblock Evercrypt: A fast, verified, cross-platform cryptographic provider.
\newblock In {\em 2020 IEEE Symposium on Security and Privacy (SP)\/} (2020),
  IEEE, pp.~983--1002.

\bibitem{Contentm50:online}
Content moderation is serious business: Pwc.
\newblock
  \url{https://www.pwc.com/us/en/industries/tmt/library/content-moderation-quest-for-truth-and-trust.html}.
\newblock (Accessed on 04/17/2022).

\bibitem{10ModelZ20:online}
10. model zoo — pytorch/serve master documentation.
\newblock \url{https://pytorch.org/serve/model_zoo.html}.
\newblock (Accessed on 10/24/2021).

\bibitem{Amazonsc3:online}
Amazon scraps secret ai recruiting tool that showed bias against women |
  reuters.
\newblock
  \url{https://www.reuters.com/article/us-amazon-com-jobs-automation-insight/amazon-scraps-secret-ai-recruiting-tool-that-showed-bias-against-women-idUSKCN1MK08G}.
\newblock (Accessed on 04/19/2022).

\bibitem{Roboflow81:online}
Roboflow: Give your software the power to see objects in images and video.
\newblock \url{https://roboflow.com/}.
\newblock (Accessed on 04/16/2022).

\bibitem{romero2021infaas}
{\sc Romero, F., Li, Q., Yadwadkar, N.~J., and Kozyrakis, C.}
\newblock Infaas: Automated model-less inference serving.
\newblock In {\em 2021 $\{$USENIX$\}$ Annual Technical Conference
  ($\{$USENIX$\}$$\{$ATC$\}$ 21)\/} (2021), pp.~397--411.

\bibitem{rotem2018glow}
{\sc Rotem, N., Fix, J., Abdulrasool, S., Catron, G., Deng, S., Dzhabarov, R.,
  Gibson, N., Hegeman, J., Lele, M., Levenstein, R., et~al.}
\newblock Glow: Graph lowering compiler techniques for neural networks.
\newblock {\em arXiv preprint arXiv:1805.00907\/} (2018).

\bibitem{russakovsky2015imagenet}
{\sc Russakovsky, O., Deng, J., Su, H., Krause, J., Satheesh, S., Ma, S.,
  Huang, Z., Karpathy, A., Khosla, A., Bernstein, M., et~al.}
\newblock Imagenet large scale visual recognition challenge.
\newblock {\em International journal of computer vision 115}, 3 (2015),
  211--252.

\bibitem{sabne2020xla}
{\sc Sabne, A.}
\newblock Xla: Compiling machine learning for peak performance, 2020.

\bibitem{shafahi2018poison}
{\sc Shafahi, A., Huang, W.~R., Najibi, M., Suciu, O., Studer, C., Dumitras,
  T., and Goldstein, T.}
\newblock Poison frogs! targeted clean-label poisoning attacks on neural
  networks.
\newblock {\em Advances in neural information processing systems 31\/} (2018).

\bibitem{simonyan2014very}
{\sc Simonyan, K., and Zisserman, A.}
\newblock Very deep convolutional networks for large-scale image recognition.
\newblock {\em arXiv preprint arXiv:1409.1556\/} (2014).

\bibitem{smutz2016tree}
{\sc Smutz, C., and Stavrou, A.}
\newblock When a tree falls: Using diversity in ensemble classifiers to
  identify evasion in malware detectors.
\newblock In {\em NDSS\/} (2016).

\bibitem{MachineL63:online}
{\sc Soni, D.}
\newblock Machine learning for content moderation — introduction.
\newblock
  \url{https://towardsdatascience.com/machine-learning-for-content-moderation-introduction-4e9353c47ae5}.
\newblock (Accessed on 04/17/2022).

\bibitem{Usergen45:online}
User-generated internet content per minute 2021 | statista.
\newblock
  \url{https://www.statista.com/statistics/195140/new-user-generated-content-uploaded-by-users-per-minute/}.
\newblock (Accessed on 04/17/2022).

\bibitem{YouTube12:online}
Youtube: hours of video uploaded every minute 2020 | statista.
\newblock
  \url{https://www.statista.com/statistics/259477/hours-of-video-uploaded-to-youtube-every-minute/}.
\newblock (Accessed on 04/17/2022).

\bibitem{StreamML22:online}
Stream.ml – machine learning, simplified!
\newblock \url{https://stream.ml/}.
\newblock (Accessed on 04/16/2022).

\bibitem{szegedy2015going}
{\sc Szegedy, C., Liu, W., Jia, Y., Sermanet, P., Reed, S., Anguelov, D.,
  Erhan, D., Vanhoucke, V., and Rabinovich, A.}
\newblock Going deeper with convolutions.
\newblock In {\em Proceedings of the IEEE conference on computer vision and
  pattern recognition\/} (2015), pp.~1--9.

\bibitem{szegedy2016rethinking}
{\sc Szegedy, C., Vanhoucke, V., Ioffe, S., Shlens, J., and Wojna, Z.}
\newblock Rethinking the inception architecture for computer vision.
\newblock In {\em Proceedings of the IEEE conference on computer vision and
  pattern recognition\/} (2016), pp.~2818--2826.

\bibitem{szegedy2013intriguing}
{\sc Szegedy, C., Zaremba, W., Sutskever, I., Bruna, J., Erhan, D., Goodfellow,
  I., and Fergus, R.}
\newblock Intriguing properties of neural networks.
\newblock {\em arXiv preprint arXiv:1312.6199\/} (2013).

\bibitem{tan2019mnasnet}
{\sc Tan, M., Chen, B., Pang, R., Vasudevan, V., Sandler, M., Howard, A., and
  Le, Q.~V.}
\newblock Mnasnet: Platform-aware neural architecture search for mobile.
\newblock In {\em Proceedings of the IEEE/CVF Conference on Computer Vision and
  Pattern Recognition\/} (2019), pp.~2820--2828.

\bibitem{httpsass80:online}
{\sc teradata}.
\newblock State of artificial intelligence for enterprises.
\newblock
  \url{https://assets.teradata.com/resourceCenter/downloads/ExecutiveBriefs/EB9867_State_of_Artificial_Intelligence_for_the_Enterprises.pdf}.
\newblock (Accessed on 04/17/2022).

\bibitem{tramer2018slalom}
{\sc Tramer, F., and Boneh, D.}
\newblock Slalom: Fast, verifiable and private execution of neural networks in
  trusted hardware.
\newblock {\em arXiv preprint arXiv:1806.03287\/} (2018).

\bibitem{tritonin57:online}
triton-inference-server/server: The triton inference server provides an
  optimized cloud and edge inferencing solution.
\newblock \url{https://github.com/triton-inference-server/server}.
\newblock (Accessed on 11/04/2021).

\bibitem{van2020lvi}
{\sc Van~Bulck, J., Moghimi, D., Schwarz, M., Lippi, M., Minkin, M., Genkin,
  D., Yarom, Y., Sunar, B., Gruss, D., and Piessens, F.}
\newblock Lvi: Hijacking transient execution through microarchitectural load
  value injection.
\newblock In {\em 2020 IEEE Symposium on Security and Privacy (SP)\/} (2020),
  IEEE, pp.~54--72.

\bibitem{volos2018graviton}
{\sc Volos, S., Vaswani, K., and Bruno, R.}
\newblock Graviton: Trusted execution environments on gpus.
\newblock In {\em 13th $\{$USENIX$\}$ Symposium on Operating Systems Design and
  Implementation ($\{$OSDI$\}$ 18)\/} (2018), pp.~681--696.

\bibitem{wan2022automated}
{\sc Wan, C., Liu, S., Xie, S., Liu, Y., Hoffmann, H., Maire, M., and Lu, S.}
\newblock Automated testing of software that uses machine learning apis, 2022.

\bibitem{wang2005translation}
{\sc Wang, Z., and Simoncelli, E.~P.}
\newblock Translation insensitive image similarity in complex wavelet domain.
\newblock In {\em Proceedings.(ICASSP'05). IEEE International Conference on
  Acoustics, Speech, and Signal Processing, 2005.\/} (2005), vol.~2, IEEE,
  pp.~ii--573.

\bibitem{AIPricin97:online}
Ai pricing | how much does artificial intelligence cost in 2020?
\newblock \url{https://www.webfx.com/martech/pricing/ai/}.
\newblock (Accessed on 04/17/2022).

\bibitem{wenzel2019literature}
{\sc Wenzel, H., Smit, D., and Sardesai, S.}
\newblock A literature review on machine learning in supply chain management.
\newblock In {\em Artificial Intelligence and Digital Transformation in Supply
  Chain Management: Innovative Approaches for Supply Chains. Proceedings of the
  Hamburg International Conference of Logistics (HICL), Vol. 27\/} (2019),
  Berlin: epubli GmbH, pp.~413--441.

\bibitem{whitehead2011precision}
{\sc Whitehead, N., and Fit-Florea, A.}
\newblock Precision \& performance: Floating point and ieee 754 compliance for
  nvidia gpus.
\newblock {\em rn (A+ B) 21}, 1 (2011), 18749--19424.

\bibitem{wuille2018libsecp256k1}
{\sc Wuille, P.}
\newblock libsecp256k1.
\newblock {\em URL: https://github.com/bitcoin/secp256k1\/} (2018).

\bibitem{xie2017aggregated}
{\sc Xie, S., Girshick, R., Doll{\'a}r, P., Tu, Z., and He, K.}
\newblock Aggregated residual transformations for deep neural networks.
\newblock In {\em Proceedings of the IEEE conference on computer vision and
  pattern recognition\/} (2017), pp.~1492--1500.

\bibitem{yin2018byzantine}
{\sc Yin, D., Chen, Y., Kannan, R., and Bartlett, P.}
\newblock Byzantine-robust distributed learning: Towards optimal statistical
  rates.
\newblock In {\em International Conference on Machine Learning\/} (2018), PMLR,
  pp.~5650--5659.

\bibitem{yin2018hotstuff}
{\sc Yin, M., Malkhi, D., Reiter, M.~K., Gueta, G.~G., and Abraham, I.}
\newblock Hotstuff: Bft consensus in the lens of blockchain.
\newblock {\em arXiv preprint arXiv:1803.05069\/} (2018).

\bibitem{yuan2019adversarial}
{\sc Yuan, X., He, P., Zhu, Q., and Li, X.}
\newblock Adversarial examples: Attacks and defenses for deep learning.
\newblock {\em IEEE Transactions on Neural Networks and Learning Systems 30}, 9
  (2019), 2805--2824.

\bibitem{zhang2020model}
{\sc Zhang, J., Elnikety, S., Zarar, S., Gupta, A., and Garg, S.}
\newblock Model-switching: Dealing with fluctuating workloads in
  machine-learning-as-a-service systems.
\newblock In {\em 12th $\{$USENIX$\}$ Workshop on Hot Topics in Cloud Computing
  (HotCloud 20)\/} (2020).

\end{thebibliography}
